\documentclass[reprint, amsmath,amssymb,aps,twocolumn, prb,groupedaddress]{revtex4-1}
\usepackage{amsfonts,amssymb}
\usepackage[subnum]{cases}
\usepackage{mathrsfs}
\usepackage{amsmath}
\usepackage{array}
\usepackage[none]{hyphenat}
\usepackage{graphicx}
\usepackage{hyperref}    
\usepackage{bm}          
\usepackage{natbib}
\usepackage{soul} 
\usepackage{color}
\usepackage{longtable}
\usepackage{ textcomp }
\usepackage{verbatim}
\usepackage{tocloft}
\usepackage[caption=false]{subfig}
\usepackage[export]{adjustbox}
\newcommand{\abs}[1]{\left| #1 \right|}
\begin{document}

\title{Roughness scattering induced insulator-metal-insulator transition in a quantum wire}
\author{Han Fu}  
\author{M. Sammon} \email{sammo017@umn.edu}  
\author{B. I. Shklovskii} 

\affiliation{Fine Theoretical Physics Institute, University of Minnesota, Minneapolis, MN 55455, USA}
\date{\today}
\begin{abstract}

We have theoretically investigated the influence of interface roughness scattering on the low temperature mobility of electrons in quantum wires when electrons fill one or many subbands. We find the Drude conductance of the wire with length $\mathcal{L}$ first increases with increasing linear concentration of electrons $\eta$ and then decreases at larger concentrations. For small radius $R$ of the wire with length $\mathcal{L}$ the peak of the conductance $G_{max}$ is below $e^2/h$ so that electrons are localized. The height of this peak grows as a large power of $R$, so that at large $R$ the conductance $G_{max}$ exceeds $e^2/h$ and a window of concentrations with delocalized states (which we call the metallic window) opens around the peak. Thus, we predict an insulator-metal-insulator transition with increasing concentration for large enough $R$. Furthermore, we show that the metallic domain can be sub-divided into three smaller domains: 1) single-subband ballistic conductor, 2) many-subband ballistic conductor 3) diffusive metal, and use our results to estimate the conductance in these domains. Finally we estimate the critical value of $R_c(\mathcal{L})$ at which the metallic window opens for a given length $\mathcal{L}$ and find it to be in reasonable agreement with experiment.
\end{abstract}

\maketitle

\maketitle

\section{Introduction}

Semiconductor nanowires attracted lots of attention due to their potential applications, such as field-effect transistors, elementary logic circuits, resonant tunneling diodes, light-emitting diodes, lasers, and biochemical sensors~\cite{Lieber,Yang}. Advances in the nanowire growth have also led to the development of novel quantum devices \cite{Lutchynrev,Hofstetter,Doh,Frolov,Fadaly}. They allow the exploration of mesoscopic transport in a highly confined system. Recently, hybrid superconductor-semiconductor nanowire devices have been identified \cite{Lutchyn,Von_oppen} as a platform to study Majorana end modes \cite{Majorana}, which exhibit topological properties \cite{Mourik,Deng_hybrid,Churchill}. To further improve this topological system, a reduction of the disorder in the nanowire is essential \cite{Sau,Onder}. Performance of some of these devices is limited by scattering of electrons on surface roughness \cite{Scheffler,Jared,Ford,Poli,Shimizu,Fengyun}. A theory of roughness scattering limited mobility of nanowires as a function of their radius $R$ and linear electron concentration $\eta$ controlled by a back gate would be helpful. In spite of some attempts to create such a theory \cite{Ando_2D,Tingwei,Sakaki_wire} the full picture of roughness limited transport in nanowires currently is not available. This is not surprising because as we show below even in the case of quantum wells there are big gaps in the roughness limited mobility theory, namely, for wells with many subbands filled. In this paper, we fill the gaps in the theoretical description of roughness limited mobility both for quantum wells and quantum wires.

Much of the focus in nanowire technology is in creating ballistic nanowires that can support the Majorana zero edge modes for quantum computation\cite{Sau,Onder}. We show below that the
possibility to achieve ballistic transport depends strongly on the radius $R$ and the length $\mathcal{L}$ of the wire. Namely, we show that for a fixed $\mathcal{L}$, there exists a critical value $R_c(\mathcal{L})$ such that electrons in wires with $R < R_c(\mathcal{L})$ are localized, while for $R > R_c(\mathcal{L})$ there is a window of concentrations where a metallic phase exists.

Before addressing why such a window exists, let us describe conventional models of roughness developed for quantum wells. In a quantum well confined by interfaces at $z=0$ and $z=L$, the surface roughness is a random shift of the interface position $\Delta(\vec{r})$ from the average level so that $<\Delta(\vec{r})>=0$, where $\vec{r}=(x,y)$ is the coordinate in $z=0$ (or $z=L$) interface plane. The roughness is described by the height correlator and its Fourier transform
\begin{equation}\label{eq:corre}
\begin{aligned}
<\Delta(\vec{r})\Delta(\vec{r'})>=&W(\vec{r}-\vec{r'}),\\
<|\Delta(q)|^2>=&W(q).
\end{aligned}
\end{equation}
First theories of surface roughness scattering have assumed the correlator to be Gaussian.\cite{Ando,Sakaki,Gold,Entin, Suris}.
\begin{equation}
\begin{aligned}
W(\vec{r}-\vec{r'})=&\Delta^2e^{-(\vec{r}-\vec{r'})^2/d^2},\\
W(q)=&\pi\Delta^2d^2e^{-q^2d^2/4}.
\end{aligned}
\end{equation}
However, experimental observations using TEM and STM measurements of Si/SiO$_2$ interfaces and InAs/GaSb interfaces found that the spacial correlations follow an exponential behavior \cite{Goodnick,Feenstra_1994}
\begin{equation}\label{eq:exp}
\begin{aligned}
W(\vec{r}-\vec{r'})=&\Delta^2e^{-\sqrt{2}|\vec{r}-\vec{r'}|/d},\\
W(q)=&\pi\Delta^2d^2(1+q^2d^2/2)^{-3/2}.
\end{aligned}
\end{equation}
This correlator describes randomly distributed flat islands of typical thickness $\Delta$ and diameter $d$ on the top of the last complete layer of the crystal~\cite{Roughness}.
On the other hand, Gaussian roughness can be visualized as randomly positioned stacks of total height $\Delta$ and diameter $d$ made of progressively smaller islands of flat atomic layers on the top of bigger ones~\cite{Roughness} similar to the ancient Mayan pyramids. As we show below, in many cases the two correlators give the same expression for the mobility in terms of $\Delta$ and $d$, and so the difference in parameter values can have serious implications. Only at very large electron densities when $k_Fd\gg1$, ($k_F$ is the Fermi wave number), do the two correlators give different expressions for the mobility. This difference is relatively unimportant for this work, so we give results only for the exponential correlator. 

While the above isotropic roughness models were designed for quantum wells with flat interfaces, they are valid for quantum wires of characteristic size $R>d$.\footnote{For the case of cylindrical wires, the characteristic size
	$R$ would be the radius of the wire. However, the results
	presented are applicable to any cross-section that can be
	described with a single characteristic length, such as a
	square wire with side length 2R or a regular hexagonal
	wire in which R is the distance from the center of the
	wire to each vertex.} In the most of this paper we deal with such roughness. However TEM images of InAs wires\cite{Fengyun} suggest that in quantum wires another model of roughness in which the radius of the wire varies along its axis may be more realistic. We discuss this ``Variable Radius Model" (VRM) and its implications in Sec. \ref{sec:FRM}.

In this paper we consider wires with linear electron concentration $\eta$ doped by a relatively distant back gate (we assume that there are no chemical donors in the wire). Then the interplay between the concentration $\eta$, the radius of the wire $R$, and the semiconductor Bohr radius $a_B$ determines the number of filled subbands of radial quantization, what is the Fermi wavenumber $k_F$ of electrons, and whether the confinement is electrostatic or by the surface barriers (referred to as geometric confinement). Here the effective Bohr radius $a_B=\kappa \hbar^2/m^* e^2$, $\kappa$ is the effective dielectric constant, $\hbar$ is the reduced Planck constant, and $m^*$ is the effective electron mass. This means that for quantum wires, there are five lengths  $\Delta$, $d$, $\eta^{-1}$, $R$, and $a_B$, or four dimensionless lengths when all are scaled by $a_B$, that determine the Drude mobility. 

\begin{figure}[h!]
	\includegraphics[width=.47\textwidth]{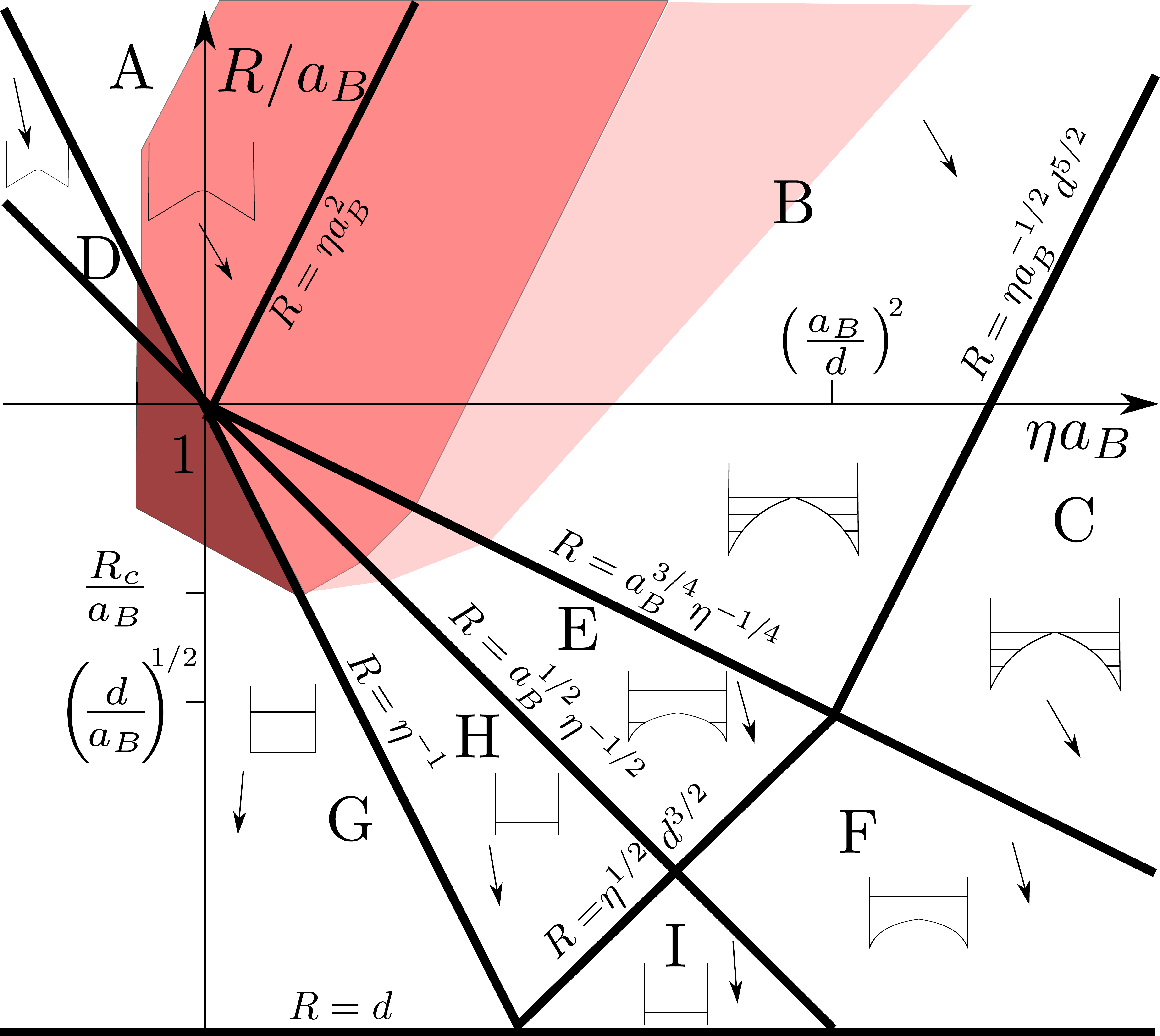}\\
	\caption{The scaling ``phase diagram" of roughness limited electron Drude mobility of a long quantum wire plotted as a function of radius $R$ and linear electron concentration $\eta$ for $d<a_B$ in the log-log scale. Different "phases" or regions are denoted by capital letters. Drude mobility expressions corresponding to these regions are given in Table \ref{tab:2}. Region boundaries are given by the equations next to them. The schematic self-consistent electron potential energy profile along the wire diameter and subbands occupied by electrons are shown for each region. Small arrows show the direction of mobility decrease in each region. The colored areas illustrate where the wire of length $\mathcal{L}$ is metallic. The dark red, light red, and pink regions correspond to the single-subband ballistic conductor, many-subband ballistic conductor, and diffusive metal regions for $R_c(\mathcal{L})< a_B$. Electrons are localized in all the colorless regions. The border between them and colored regions is determined by the length of the wire $\mathcal{L}$. We assumed that $\mathcal{L}\sim 1$ $\mu$m as is used in quantum devices. For shorter wires $R_c(\mathcal{L})$ decreases, and the colored metallic regions expand to cover most of the area of the phase diagram.}\label{fig:wire}
\end{figure}

Below we use the scaling theory to calculate the low temperature roughness limited Drude mobility $\mu$ in units $\left(e/\hbar\right)\left(d^4/\Delta^2\right)$ as a function of the dimensionless variables $R/a_B$ and $\eta a_B$. Here the use of Drude's name signifies that we ignore interference effects and electron-electron correlations. We summarize our results for different regions in Fig. \ref{fig:wire} as a ``phase diagram" in the plane of $R/a_B$ and $\eta a_B$, the details of which are elaborated in Sec. \ref{sec:wire}. For the most interesting case $\Delta\ll d\ll a_B$ we find a total of 9 regions $A-I$ whose mobilities are listed in Table \ref{tab:2}. It should be noted that due to the limitations of the scaling theory, the mobility expressions for the different regions of the phase diagram are valid only away from the borders between different regions. In the vicinity of the border between regions, there is a smooth crossover between the two mobilities, the details of which are beyond the scope of this paper. While the scaling approach only gives the dependence of mobility on the different parameters without numerical precision, its simplicity allows for a clear picture of the different physical domains and the approximate limits under which they occur. 

\begin{figure}[h]
	\includegraphics[width=.47\textwidth]{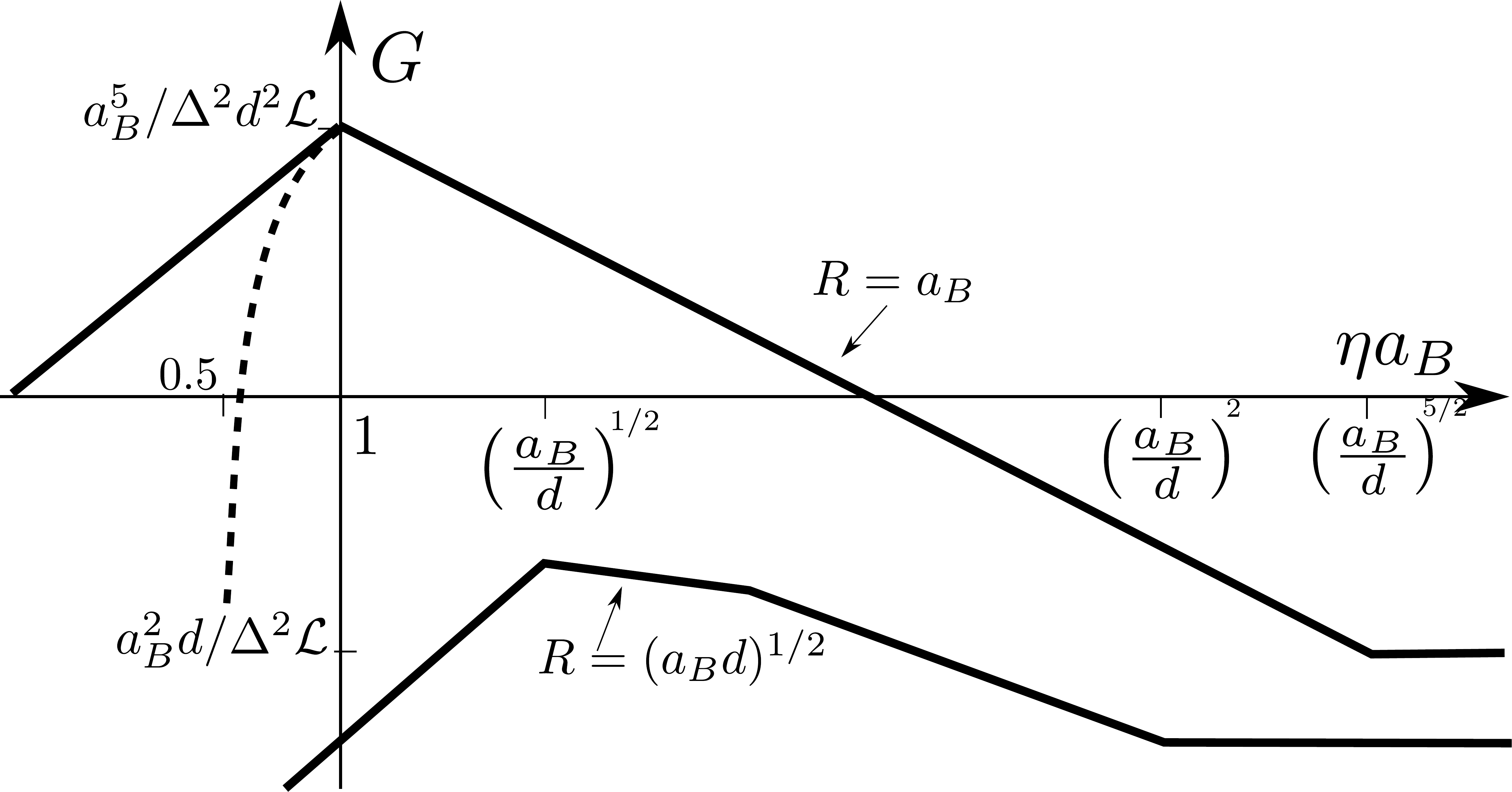}\\
	\caption{The scaling behavior of the dimensionless Drude conductance of a quantum wire with length $\mathcal{L}$ and radius $R$ as a function of the linear electron concentration $\eta$ at different wire radii for $d\ll a_B$ in the log-log scale (full lines). The upper curve corresponds to $R=a_B$ and the lower one is $R=\left(a_Bd\right)^{1/2}$. They are obtained from cross-sections of the ``phase diagram" in Fig. \ref{fig:wire} and the mobilities in Table \ref{tab:2}. The dashed line on the upper curve shows the metal-insulator crossover near $\eta a_B = 0.5$ induced by electron-electron interactions. We see that for $R=a_B$ the metallic window is open, while for $R=(a_Bd)^{1/2}$ the window is closed. $\mathcal{L}=a_B^{7/2}\Delta^{-2}d^{-1/2}$ was chosen. }\label{fig:conductance}
\end{figure}

Now we are ready to address the origin of the Drude conductance peak which leads to a metallic window for large $R$, illustrated by the colored regions in Fig. \ref{fig:wire}. Schematic plots of the Drude conductance (in units $e^2/h$) $G=\eta\mu h/\mathcal{L}e$ of the wire with length $\mathcal{L}$ are shown in Fig. \ref{fig:conductance} for two representative values of $R$ by full lines. They are obtained from cross sections of Fig. \ref{fig:wire} and the mobilities in Table \ref{tab:2}. At low concentrations, we see that the Drude conductance increases with increasing concentration. This corresponds to Region G of Fig. \ref{fig:wire}, where there is a single radial subband occupied and the electrons are confined geometrically. We know from Fermi's golden rule that the relaxation time $\tau$ is inversely proportional to the density of states at the Fermi energy, which in the one-dimensional (1D) case goes like $1/k_F\sim1/\eta$. The scattering potential however is independent of concentration in this regime.  Therefore, the relaxation time $\tau$, the mobility $\mu$, and conductance $G$ increase with concentration due to the decrease in the density of states. This trend continues until the concentration is large enough that multiple subbands become occupied. Now electrons have more states to scatter into, and the relaxation time quickly decreases with increasing concentration. Thus the conductance peaks at the border concentration $\eta_c$ when electrons begin to populate multiple subbands. The peak of the Drude conductance for the most interesting cases of $R\leq a_B$ is given by 
\begin{equation}\label{eq:conductancepeak}
G_{max}=\frac{R^5}{\mathcal{L}\Delta^2d^2}.
\end{equation}

So far we have ignored electron-electron interactions and quantum interference effects. 
They dramatically change the conductivity of one dimensional systems at low temperatures.
For single subband wires (regions D and G) electron-electron interactions result in Wigner-crystal-like correlations and pinning of the electron gas leading to the metal-insulator crossover near $\eta a_B =0.5$\cite{SRG}. In Fig. \ref{fig:conductance} the corresponding collapse of conductance at $\eta a_B < 0.5$ is shown by the dashed lines. According to Luttinger liquid theory~\cite{KF,FG} similar effects persist at very low temperatures in very long wires. We are interested here in relatively  short wires with $\mathcal{L} \sim 1$ $\mu$m, where plasmon quantization does not allow such effects to develop~\cite{FG}. Therefore, for $\eta a_B > 1$ we can ignore electron-electron interactions. However, in this case we should still take into account quantum interference effects. They lead to one-electron localization when Drude $G < 1$. This means that when $G_{max} < 1$ (see lower curve in Fig. \ref{fig:conductance}), the wire is an insulator at any concentration $\eta$. On the other hand, for $G_{max} >  1$ (see upper curve in in Fig. \ref{fig:conductance}) the wire has a concentration window of metallic behavior. The critical radius $R_c(\mathcal{L})$ in which the metallic window opens is then determined by the condition that $G_{max}=1$. For $G_{max}$ defined by Eq. (\ref{eq:conductancepeak}) we find 
\begin{equation}\label{eq:Rc}
R_c(\mathcal{L})=(\Delta^2d^2\mathcal{L})^{1/5}.
\end{equation}
Note that the restriction that $\eta a_B>0.5$ necessary for the single subband wires to be metallic requires $R_c<2 a_B$. 

Thus, we predict a zero temperature reentrant insulator-metal-insulator transition with increasing $\eta$ in quantum wires with $R > R_c(\mathcal{L})$.
Such a transition was first predicted for a two-dimensional electron gas (2DEG) in silicon MOSFET\cite{Sarma}. However it was later shown\cite{Roughness} that there is no second reentrant metal-insulator transition at large concentrations of a 2DEG as the dimensionless conductance saturates at a value larger than unity. As our paper shows the idea of Ref. \onlinecite{Sarma} is realized in quantum wires. (For more details see our Sec. II below.)

The metallic regimes for a wire with $R_c(\mathcal{L})< a_B $ are shown in different colors in Fig. \ref{fig:wire}, while regions where the electrons are localized are left blank. The dark red, light red, and pink colored regions of the metallic regime specify a single-subband ballistic conductor, a many-subband ballistic conductor, and a diffusive metal respectively. It should be emphasized that the metal-insulator and ballistic-diffusive borders depend on the wire length. With decreasing $\mathcal{L}$  and $R_c(\mathcal{L})$ the colored regions expand dramatically and for short wires eventually cover most of the phase diagram. In Fig. 1 we used $\mathcal{L}\sim1$ $\mu$m as in Fig. 2, which is typically used in quantum devices (see details in Sec. \ref{sec:disc}). 

 The detailed derivation of all the metallic border equations are given in Sec. \ref{sec:conductance} and in Tab. \ref{tab:border}.
 Here we give a brief summary of the derivation. Let us begin with the metal-insulator border. For $\eta a_B>0.5$ this border comes from the condition that the Drude conductance $G_D=e^2/h$, and gives rise to the sequence of border lines between the colored and uncolored regions with minimum at $R_c(\mathcal{L})$ in Fig. \ref{fig:wire}. For $\eta a_B<0.5$ there is no metallic regime for the single subband regions (G and H), as illustrated by the vertical line that cuts off the dark red region of Fig. \ref{fig:wire} at low concentrations. This line continues vertically to the asymptotic line $\eta a_B\sim C (R/a_B)$, which can be understood as the Wigner crystallization of the 2DEG at $na_B^2=C\ll1$, where $n=\eta/2\pi R$. Finally we address the ballistic-diffusive border which only exists in the regions with many subbands occupied. Typically, a diffusive metal becomes ballistic when the mean free path $l=\mathcal{L}$. However, for the many subband regions there is an ambiguity, as we can have different $l$ for different subbands. Fortunately, the conductance in these cases is determined by a small subset of subbands which have identical $l$ and we define the ballistic-diffusive border by the line where $l=\mathcal{L}$ for these subbands.

Let us discuss the conductance in the different colored regions of Fig. 1. We begin with the ballistic regimes (red regions of Fig. \ref{fig:wire}). Here the dimensionless conductance $G\approx2K$, where $K$ is the number of ballistic channels of a wire with finite length $\mathcal{L}$, and the factor of 2 comes from the spin degeneracy. Estimates of $K$ can be found in Sec. \ref{sec:conductance}. Within the diffusive regime (pink regions of Fig. \ref{fig:wire}) $G=(h/e)\eta\mu/\mathcal{L}$, where the mobility is given in Tab. \ref{tab:2}. Finally, in the insulating regions electrons are localized at temperature $T=0$. At finite $T$ wires conduct via phonon assisted hopping. 
Calculations of the hopping conductivity are relatively straightforward, but are beyond the scope of this paper.

The plan of this paper is as follows. In Secs. \ref{subsec:result_well} and \ref{subsec:physical_well} we study the roughness limited mobility of quantum wells as a function of their width $L$ and two-dimensional concentration of electrons $n$ arriving at the ``phase diagram" for $\mu (L,n)$ with nine different regions.  In Sec. \ref{sec:wire} we use the quantum  well ``phase diagram"  to construct the ``phase diagram" $\mu(R,\eta)$ for quantum wires with surface roughness described by Eq. \eqref{eq:exp}. In Sec. \ref{sec:conductance} we use our results for the Drude mobility to estimate the wire conductance in the ballistic and diffusive regions. In Sec. \ref{sec:FRM} we discuss quantum wires within the Variable Radius Model (VRM). In Sec. \ref{sec:disc} we dwell upon some experimental implications, namely the peak mobility and the value of radius $R_c(\mathcal{L})$ in which the metallic window opens.  We conclude with Sec. \ref{sec:con}.

\section{Roughness limited mobility results for quantum wells}
\label{subsec:result_well}

\begin{figure}[h]
\includegraphics[width=.46\textwidth]{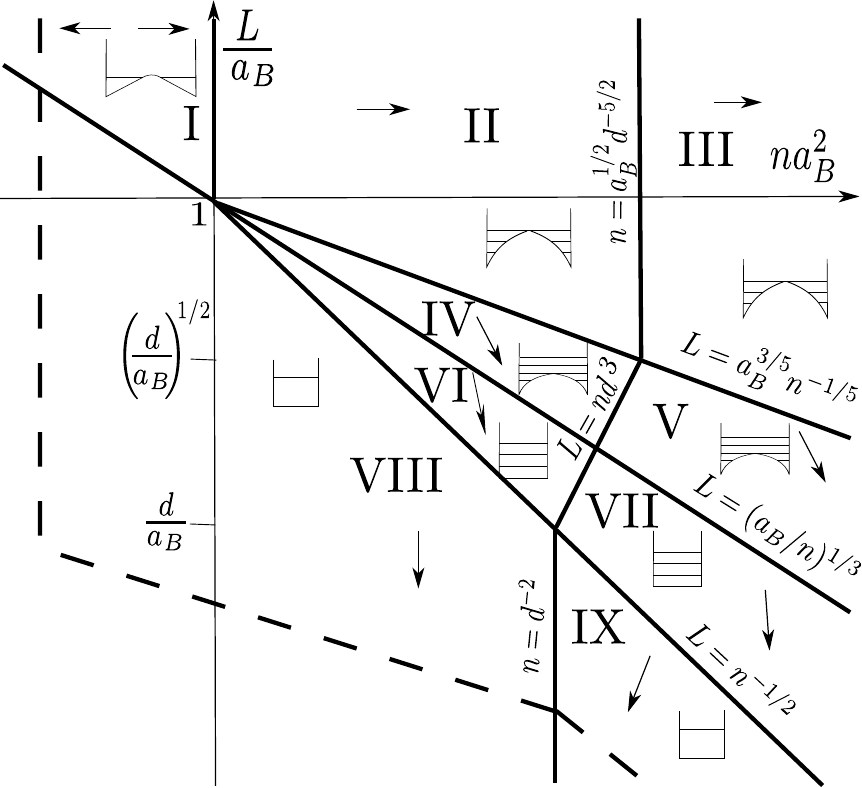}\\
\caption{The scaling "phase diagram" of roughness limited electron mobility of quantum well at different well width $L$ and 2D electron concentration $n$ for $d\ll a_B$ in the log-log scale. Different "phases" or regions are denoted by Roman numerals. Mobility expressions corresponding to these regions are given in Table \ref{tab:1}. Region boundaries are given by the equations next to them. The schematic self-consistent electron potential energy profile along the $z$-axis of wells and  levels (subbands) occupied by electrons are shown for each region. Small arrows show the direction of mobility decrease in each region. Apparently the maximum mobility is achieved in Region I. The dashed line indicates schematically the border of the metal-insulator transition (MIT) at small enough $n$. At large $n$ there is no reentrant MIT in spite of the decreasing mobility.}\label{fig:well}
\end{figure}

To understand the roughness limited mobility of quantum wires, it is convenient to first make clear of that in quantum wells. We start from a quantum well confined by two high potential barriers at $z=0,\, L$. It has the two-dimensional (2D) electron concentration $n$ created either by two positive donor layers located symmetrically on both sides of the well or by two symmetric metallic gates. In both cases, at $z = 0,\,L$ there is an electric field pointing into the well with $|E|= 2\pi n e$, where $e$ is the electron charge. Interplay of effects of the electric field and barrier confinement creates 5 different types of wells shown in Fig. \ref{fig:well} in regions I - IX. In a narrow well the electric field $E$ plays a minor role in level quantization compared to confining barriers so that we assume that all subbands are geometrically confined in the small $L$ regions VI, VII, VIII, and IX  in Fig. \ref{fig:well}. When the concentration is relatively small, electrons occupy only the first subband. At larger $n$ electrons populate many subbands (see the level schematics in regions VI and VII in Fig. \ref{fig:well}). In wider wells shown in regions I, II, III, IV and V the electric field becomes important compared to the surface barriers. In turn this leads to the splitting of the electron density in two peaks. With growing $L$, in the beginning (regions IV and V) this splitting is moderate and affects only the lowest subbands. In regions II and III the splitting results in two separate accumulation layers in response to the electric field each side of the well. Finally at large $L$ and small $n$ we again reach the single subband limit, however the confinement is electrostatic rather than geometric (region I  in  Fig. \ref{fig:well}).

The roughness limited mobility of a single-subband electron gas of a quantum well (regions I, VIII and IX) was thoroughly studied in Refs. \onlinecite{Ando,Sakaki,Gold,Entin} more than 30 years ago. On the other hand, the roughness limited mobility of accumulation layers was calculated recently in Ref. \onlinecite{Roughness}, results of which are directly applicable to regions II and III. However, no work has been done in the intermediate regions where many subbands are occupied but the electric field is weak so that some or all of the subbands are confined geometrically (regions IV, V, VI, and VII). In this paper we fill this gap. Below, because of the complexity of the problem, we first present the final results in this section and then give their derivations in next section.

The complete results at $d\ll a_B$ are shown in Fig. \ref{fig:well} and Table \ref{tab:1}. The single subband results I, VIII, and IX are taken from Refs. \onlinecite{Ando,Sakaki,Gold,Entin} and accumulation layer results II and III are from Ref. \onlinecite{Roughness}. For the intermediate regions IV, V, VI, and VII, we obtain their results in this paper.

Let us first look at the physical meaning and corresponding equations of boundary lines in Fig. \ref{fig:well}. Across the line between Region I and Region II, the concentration becomes so large that electrons have to occupy multiple subbands (see level schematics in Fig. \ref{fig:well}). With $n$ further increased, $k_Fd$ becomes larger than unity in Region III where $k_F$ is the three-dimensional (3D) electron Fermi wavenumber here. Instead of averaging over different islands, the electron hits only a single island now. This leads to the change of the mobility result at the II-III border.

For regions I, II, and III, all subbands are electrostatically confined. For moderately smaller well width $L$, some of the subbands become geometrically confined. This happens when the well width $L$ becomes smaller than the characteristic thickness $D$ of the accumulation layer, where \cite{Frenkel,Schecter}
\begin{equation}
D\simeq\frac{a_B}{\left(na_B^2\right)^{1/5}}.\label{eq:decay_length}
\end{equation}
The criterion $L=D$ then gives the line between II, III and IV, V. At the line between IV and V, $k_Fd=1$.

\begin{table}[h]
\caption{\label{tab:1} Mobility $\mu$ in units of $\left(e/\hbar\right)\left(d^4/\Delta^2\right)$ as a function of the 2D electron concentration $n$ at $d<a_B$ for different regions.}
\begin{ruledtabular}
\renewcommand{\arraystretch}{2}
\begin{tabular}{c|c|c}
 I&II &III\\ \hline
 $a_B^2/n^2d^6$ &$a_B^{8/5}/n^{11/5}d^6\quad\quad$&$a_B/nd^3$\\ \hline
  IV & V & VI\\ \hline
 $a_B^{1/2}L^{11/6}/n^{11/6}d^6$& $a_B^{1/2}L^{5/6}/n^{5/6}d^3\quad$&$L^{10/3}/n^{4/3}d^6$\\ \hline
 VII & VIII & IX\\ \hline
 $L^{7/3}/n^{1/3}d^3$ &$L^6/d^6\quad$&$L^6n^{3/2}/d^3$\\
\end{tabular}
\end{ruledtabular}
\end{table}

With further reduction of $L$, all subbands would be geometrically confined (see the level schematic in Fig. \ref{fig:well}). This happens when the electrostatically confined distance of the lowest subband electrons from the surface is equal to the well width $L$. This distance is $D_0 \simeq a_B^{1/3}/n^{1/3}$ (see Refs. \onlinecite{Ando, Entin}) which is the smallest among all subbands since the lowest subband has the smallest kinetic energy in the $z$ direction. The condition $L=D_0$ gives the line between IV, V and VI, VII. The border between VI and VII corresponding to the critical point of $k_Fd=1$ is a continuation of the line between regions IV and V. Moving to even smaller $L$ from regions VI and VII, we cross over to the single subband (see the level schematic in Fig. \ref{fig:well}). This corresponds to the line $k_FL=1$ between VI, VII and VIII, IX. The border of the VIII and IX regions is the line of $k_Fd=1$ where $k_F$ is the 2D electron Fermi wavenumber here. In Fig. \ref{fig:well}, one can see that there is another border line between I and VIII, which both correspond to a single subband gas. However, Region  I corresponds to two electrostatically split electron subbands near the two well interfaces, while Region VIII represents the case that the electron subband is spatially restricted by the well width $L$ (see the level schematic in Fig. \ref{fig:well}). Their crossover happens at the point that both electrostatic and geometric confinements give the same thickness of the electron gas. Remember that the electrostatically confined thickness of the first subband is $D_0$. Then the condition $L=D_0$ gives the border. So this line between I and VIII is just an extension of the line between IV, V and VI, VII.

One should note that here in Table \ref{tab:1}, all results are shown without numerical coefficients, i.e., we present only the scaling behavior. Previous works have already found the exact coefficients in the single subband regions I, VIII, and IX \cite{Ando,Sakaki}. Results of many subband regions II, IV, and VI with $k_Fd\ll1$ can also be obtained with the approximate coefficients as seen later in Sec. \ref{subsec:physical_well}. We cannot get coefficients analytically in remaining regions III, V, and VII. Thus we focus only on the scaling behaviors in all tables and derivations.

In Fig. \ref{fig:well} the metal-insulator transtition (MIT) is shown schematically by the dashed lines. Let us dwell on the meaning of these lines. The lower line is related to the localization physics of a non-interacting electron gas. Strictly speaking all states are localized in 2D infinite samples, however at $k_Fl\gg1$ the localization length grows exponentially as $\zeta=l\exp ( k_F l)$, where $l$ is the mean free path. In finite square samples of area $A$ we have in mind that $\zeta$ quickly becomes larger than the sample size $A^{1/2}$. This allows one to discuss the metallic conductivity and expect the insulator-metal transition near $\sigma=(e^2/\hbar)\ln(A^{1/2}/l)$. Ignoring the logarithm and using the expressions of mobility $\mu$ for VIII and IX in Table \ref{tab:1} as well as $\sigma=ne\mu$, one gets that the low $L$ MIT border of Region VIII obeys $L=\Delta^{1/3} d^{1/3}n^{-1/6}$. We also find the MIT border of Region IX is $L=\Delta^{1/3}d^{-1/6}n^{-5/12}$. We have used $\Delta/d=\left(d/a_B\right)^{8/5}$ in order to draw these lines. The vertical line $na_B^2=C\ll1$ reflects the role of the Coulomb interaction between electrons in a degenerate electron gas. At $na_B^2\ll1$ strong Coulomb repulsion leads to Wigner crystallization. The Wigner crystal is pinned by relatively small disorder and electrons become localized.

\section{Roughness limited mobility derivations for quantum wells}
\label{subsec:physical_well}

In the previous section, we have presented the physical picture of all 9 regions and their border lines and summarized the mobility results. In this section, we derive the new expressions of mobility for regions IV, V, VI, and VII.
First, let us derive $\mu$ for Region VI. According to Fermi's golden rule and the Boltzmann equation, the relaxation time $\tau_N$ of a particular state with the wavefunction $\xi(z,\vec{r})$ and with in-plane velocity $\vec{v_k}$ in the $N$-th (counted from the bottom lowest subband) subband is
\begin{equation}\label{eq:scattering_rate}
\frac{1}{\tau_N}=\frac{2\pi}{\hbar} \sum_{N'}\int\frac{d^2k'}{(2\pi)^2} \frac{\abs{V(q)}^2}{\epsilon(q)^2}\delta(\varepsilon-\varepsilon_F)\left(1-\frac{\vec{v_{k'}}\cdot\vec{E}}{\vec{v_k}\cdot\vec{E}}\frac{\tau_N'}{\tau_N}\right),
\end{equation}
where $\tau_N,\,\tau_N'$ denote the relaxation time for $N,\,N'$-th subbands, $\vec{v}_{k'}$ is the in-plane velocity for the final state with the wavefunction $\xi'$ in the $N'$-subband with in-plane momentum $\vec{k'}$, $\varepsilon$ is the energy of the final state $\xi'$ and $\varepsilon_F$ is the Fermi energy, $q$ is the transferred momentum in the $x-y$ plane between $\xi$ and $\xi'$. Here $V$ is the scattering matrix element arising from the scattering potential. Due to the electronic screening, the Fourier transform of the scattering potential $V(q)$ is reduced by the dielectric function $\epsilon(q)$ \cite{Ando}. One should note that here the last term inside the parenthesis of Eq. (\ref{eq:scattering_rate}) does not reduce to $\cos\theta$, where $\theta$ is the angle between initial and final total momenta. This is because, due to the 2D nature of the surface roughness and thus of the scattering potential, the multisubband electron gas experiences anisotropic scattering, i.e., different subbands have different relaxation times \footnote{One should note here that the definition of relaxation time $\tau$ is still valid according to Ref. \onlinecite{Siggia}, which might be broken in more complicated cases.}. As a result $\cos\theta$ in Eq. (\ref{eq:scattering_rate}) is replaced by the ratio of the out-of-equilibrium part of distribution function of the states $\xi'$ and $\xi$ represented by $\left(\vec{v_{k'}}\cdot\vec{E}/\vec{v_k}\cdot\vec{E}\right)\left(\tau_N'/\tau_N\right)$ (see Ref. \onlinecite{Siggia}). For brevity, we refer to this term as the distribution function ratio (DFR) from now on.

It is known that the roughness-caused scattering potential $V(\vec{r})$ and corresponding scattering matrix element $V(q)$ satisfy the equation \cite{Ando,Roughness}
\begin{equation}\label{eq:scattering_matrix element}
\begin{aligned}
V(r)=\frac{\hbar^2}{m^*}\Delta(\vec{r})\left.\frac{\partial \xi}{\partial z}\frac{\partial \xi'}{\partial z}\right\vert_{z=0, L},\\
<|V(q)|^2>\simeq\left(\frac{\hbar^2}{m^*}\right)^2\frac{k_z^2}{Z}\frac{k_z'^2}{Z'}W(q),
\end{aligned}
\end{equation}
where $k_z\simeq N/Z,\,k_z'\simeq N'/Z'$ are the $z$-direction momenta of $\xi$ and $\xi'$, $Z$ and $Z'$ are the $z$-direction widths of the $ N$-th and $N'$-th subbands, which are determined by the confinement. For example, when the subband $N$ is electrostatically confined, $Z=\varepsilon_z/eE\simeq\hbar^2k_z^2/m^*e^2n\simeq a_Bk_z^2/n$ ($\varepsilon_z$ is the kinetic energy in $z$-direction), while when geometrically confined, $Z=L$. For Region VI, all subbands are geometrically confined. So
\begin{equation}
<|V(q)|^2>\simeq\left(\frac{\hbar^2}{m^*}\right)^2\frac{N^2}{L^3}\frac{N'^2}{L^3}W(q).
\end{equation}
Since in Region VI $k_Fd\ll 1$, $W(q)\simeq \Delta^2 d^2$ is independent of $q$ according to Eq. \eqref{eq:exp}. The scattering is isotropic for a given subband $N'$ with respect to different directions of $\vec{v}_{k'}$. The scattering rate is then reduced to
\begin{equation}\label{eq:scattering_rate_small_d}
\frac{1}{\tau_N}=\frac{2\pi}{\hbar} \sum_{N'}\int\frac{d^2k'}{(2\pi)^2} \left(\frac{\hbar^2}{m^*}\right)^2\frac{N^2N'^2\Delta^2d^2}{L^6\epsilon(q)^2}\delta(\varepsilon-\varepsilon_F).
\end{equation}
The (2D) screening radius is $a_B/k_FL$ where $k_FL$ is the total number of subbands in Region VI. Since $L\ll a_B$ in this region, this screening radius is much larger than the Fermi wavelength $1/k_F$. So $\varepsilon(q)\approx 1$ and the screening can be ignored for the scattering between $N$-th subband and the typical subbands with $k_z'\simeq k_F$ and thus $q\sim k_F$.
Eq. \eqref{eq:scattering_rate_small_d} then yields
\begin{equation}\label{eq:scattering_rate_VI}
\begin{aligned}
\frac{1}{\tau_N}\simeq\frac{\hbar}{m^*}\frac{N^2\Delta^2d^2}{L^6} \sum_{N'}N'^2
\simeq \frac{\hbar}{m^*}\frac{N^2\Delta^2d^2}{L^6}\left(k_FL\right)^3\,\\
\simeq \frac{\hbar}{m^*}\frac{N^2\Delta^2d^2k_F^3}{L^3}\quad \quad,\\
\end{aligned}
\end{equation}
where the 3D wavenumber $k_F=\left(n/L\right)^{1/3}$, and the scattering rate is mainly determined by scattering between the $N$-th subband and typical subbands with large $N'$. The absence of screening in the scattering rate calculation is then self-consistently justified. Also, from Eq. \eqref{eq:scattering_rate_VI}, one can easily see that $\tau_N\propto 1/N^2$ so the lowest subband with $N=1$ has the largest relaxation time while for typical subbands with $k_z\simeq k_F$ and, thus, $N\simeq k_FL$, the corresponding relaxation time is $\left(k_FL\right)^2$ times smaller. Since there are $\sim k_FL$ subbands in total with each subband having a 2D concentration $n/k_FL$ and the number of typical subbands is close to the total number $k_FL$, the final conductivity is dominated by the lowest subband as
\begin{equation}\label{eq:conductivity_VI}
\sigma=\frac{n}{k_FL}\frac{e^2}{\hbar} \frac{L^3}{\Delta^2d^2k_F^3}=ne\frac{e}{\hbar}\frac{L^2}{\Delta^2d^2k_F^{4}},
\end{equation}
and the effective mobility is
\begin{equation}\label{eq:mobility_VI}
\mu=\frac{\sigma}{ne}=\frac{e}{\hbar}\frac{L^2}{\Delta^2d^2k_F^{4}}=\frac{e}{\hbar}\left(\frac{d^4}{\Delta^2}\right)\frac{L^{10/3}}{d^6n^{4/3}}.
\end{equation}
This is the result shown in Table \ref{tab:1} in Sec. \ref{subsec:result_well}.

Now let us move to Region IV. This region is a crossover between completely geometrically confined Region VI to completely electrostatically confined Region II. The lowest $M$ subbands are electrostatically confined due to their relatively small distances to the surface while the $k_FL-M$ higher subbands are geometrically confined. So for the lowest $M$ subbands, $k_z^2/Z\sim n/a_B$ is a constant independent of the subband index $N$ determined only by the surface electric field $E$ or the 2D electron concentration $n$. As a result, the lowest $M$ subbands have comparable relaxation times. The rest $k_FL-M$ subbands are geometrically confined and their contribution to the conductivity is dominated by the lowest subband of the group, i.e., by the $(M+1)$-th subband. Here the index $M$ is obtained by the condition that its electrostatically confined width is equal to the well width $L$
\begin{equation}
\begin{aligned}
\frac{a_Bk_z^2}{n}=L,\quad\quad
k_z\simeq\frac{M}{L}.
\end{aligned}
\end{equation}
As a result, $M=\left(nL^3/a_B\right)^{1/2}$. Now Eq. \eqref{eq:scattering_rate_VI} is modified for subbands from 1 to $M$ as
\begin{equation}\label{eq:scattering_rate_IV}
\begin{aligned}
\frac{1}{\tau_{1-M}}\simeq&\frac{\hbar}{m^*}\frac{n\Delta^2d^2}{a_B} \left(\sum_{N'=1,\dots,M}\frac{n}{a_B}
+\sum_{N'=M+1,\dots, k_FL}\frac{N'^2}{L^3}\right)\\
\simeq& \frac{\hbar}{m^*}\frac{n\Delta^2d^2}{a_B}\sum_{N'=M+1,\dots,k_FL}\frac{N'^2}{L^3}
\simeq \frac{\hbar}{m^*}\frac{n\Delta^2d^2k_F^3}{a_B},\quad \quad\\
\end{aligned}
\end{equation}
where $N'^2/L^3\gg n/a_B$ for $N'>M$ and the total number of subbands is still $k_FL\gg M$ in Region IV. Therefore the scattering rate of each subband is always determined by its scattering into the typical subbands which are geometrically confined to a width $L$ and have the momentum $k_z=k_F$ in the $z$-direction. One can easily check that in Region IV, i.e., at $L<n^{-1/5}a_B^{3/5}$, the conductivity is determined by the lowest $M$ subbands and the effective mobility
\begin{equation}\label{eq:mobility_IV}
\begin{aligned}
\mu=&\frac{\sigma}{ne}=\left(M\times \frac{n}{k_FL} \frac{e^2}{m^*} \frac{m^*}{\hbar}\frac{a_B}{n\Delta^2d^2k_F^3}\right)\frac{1}{ne}\\=&
\frac{e}{\hbar}\left(\frac{d^4}{\Delta^2}\right)\frac{a_B^{1/2}L^{11/6}}{n^{11/6}d^6}
\end{aligned}
\end{equation}
is obtained in a way similar to that of Region VI discussed before. This is the result given in Table \ref{tab:1}.

Now let us talk about the $k_Fd\gg 1$ case for regions V and VII. In this case, $W(q)$ is no longer a constant but can be much smaller than $\Delta^2d^2$ for some values of $q$. The scattering is no longer isotropic in the $x-y$ plane and one cannot ignore the DFR term $\vec{v_{k'}}\cdot\vec{E}\tau_N'/\vec{v_k}\cdot\vec{E}\tau_N$ in Eq. \eqref{eq:scattering_rate}. As we show in Appendix \ref{App:AppendixA}, the scattering is dominated by events with $q\simeq k_F$ instead of small $q\lesssim 1/d$. It can be easily seen quasi-classically that only when an electron hits the sharp edge of an island can the non-specular reflection happen. This is an event on a length scale $k_F^{-1}\ll d$ so that the scattering is dominated by $q\simeq k_F$.

For the dominant large angle scattering, though the term $(1-\vec{v_{k'}}\cdot\vec{E}\tau_N'/\vec{v_k}\cdot\vec{E}\tau_N)$ after averaging over different $\phi$ is not exactly unity like in the $k_Fd\ll 1$ case, it is still of order unity. Thus in the scaling sense, the difference brought by $k_Fd\gg1$ is only in the $(k_Fd)^3$ times reduction of $W(q)$. (One should note that for the large angle scattering, the rate is dominated by scattering into typical subbands of $k_z'\simeq k_F$ similarly to the $k_Fd\ll 1$ case discussed before. The screening here is again ignored since the large angle scattering has $q\simeq k_F$ and the screening radius $a_B/k_FL$ is much larger than the electron Fermi wavelength $k_F^{-1}$, similarly to the case in regions IV and VI.) As a result, from Region IV to V, the scattering rate decreases by $(k_Fd)^3$ for each subband and the effective mobility increases by $(k_Fd)^3$. A similar increase by a factor $(k_Fd)^3$ happens across the border from Region VI to Region VII. So far we have derived all the new results in Fig. \ref{fig:well} and Table \ref{tab:1}.

One can see from Fig. \ref{fig:well} and Table \ref{tab:1} that the results of mobility in different regions match each other at all borders between the regions. Actually, using the derived result Eq. \eqref{eq:mobility_VI} for Region VI together with the results for regions II, III, and IX, which are already known, one can uniquely identify the mobility expressions in regions IV, V, and VII by matching them with the neighboring mobilities on the borders.

So far, we have been focused on the $d\ll a_B$ case, which is generic for large $a_B$ semiconductors such as InAs and InSb. Now we would like to briefly discuss the $d \gg a_B$ case, which may take place, say, in silicon. Let us start from the case when $d= a_B$. In this case, the phase diagram Fig. \ref{fig:well} is dramatically simplified as the middle regions II, IV, and VI vanish and the border line $k_Fd=1$ merges with the vertical axis $na_B^2=1$. Let us now move to the case $d\gg a_B$. Since at $na_B^2 \ll 1$, the electron gas is two-dimensional for all values of $L$, there is only one line $nd^2=1$ for the critical border $k_Fd=1$. We assume that this line is located already in the insulator regime, so that in the whole metallic region $k_Fd \gg 1$. This leads to an additional factor $(k_Fd)^3$ to the mobility result in Region I and gives $\mu=\left(e/\hbar\right)\left(d^4/\Delta^2\right)\left(a_B^2/n^{1/2}d^3\right)$ (see Ref. \onlinecite{Roughness}). Mobility results for the extended regions III, V, VII, and IX remain the same as in Table \ref{tab:1}.

\section{roughness limited mobility in quantum wires}
\label{sec:wire}

In the previous sections, we described the roughness limited mobility in a quantum well as a function of the 2D electron concentration $n$ and the well width $L$. Here we would like to generalize these results to that of a nanowire with linear electron concentration $\eta$ and radius $R$.  We assume that an electric field $E=2e\eta/R$ applied radially inward at surface of the wire. Such a system can be realized by a metallic gate surrounding the nanowire, or a planar gate located a distance larger than the wire radius $R$.

\begin{figure}[h!]
	\subfloat{(a)
		\centering
		\includegraphics[width=.13\textwidth,valign=t]{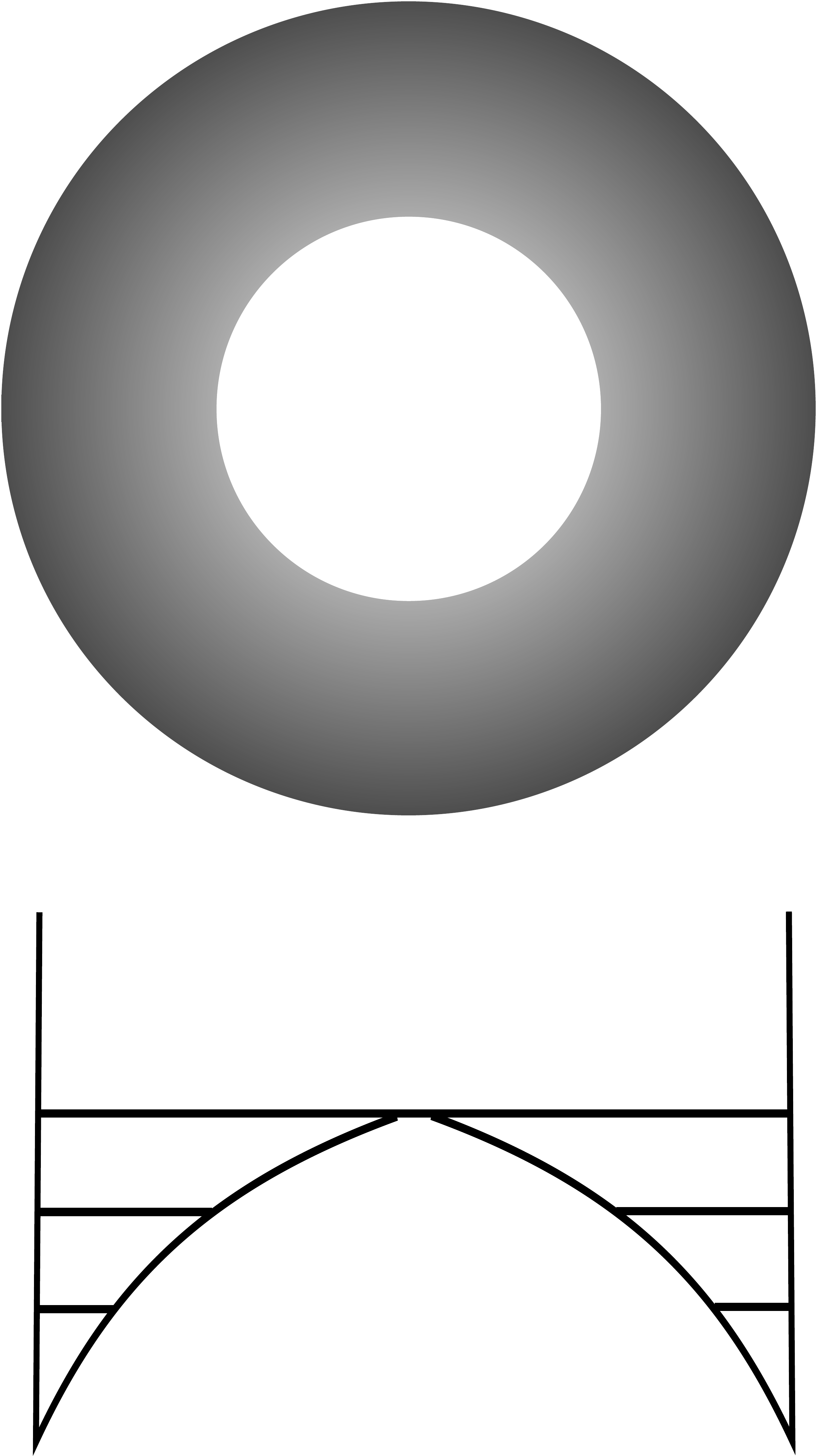}
	}
	\subfloat{(b)\includegraphics[width=.13\textwidth,valign=t]{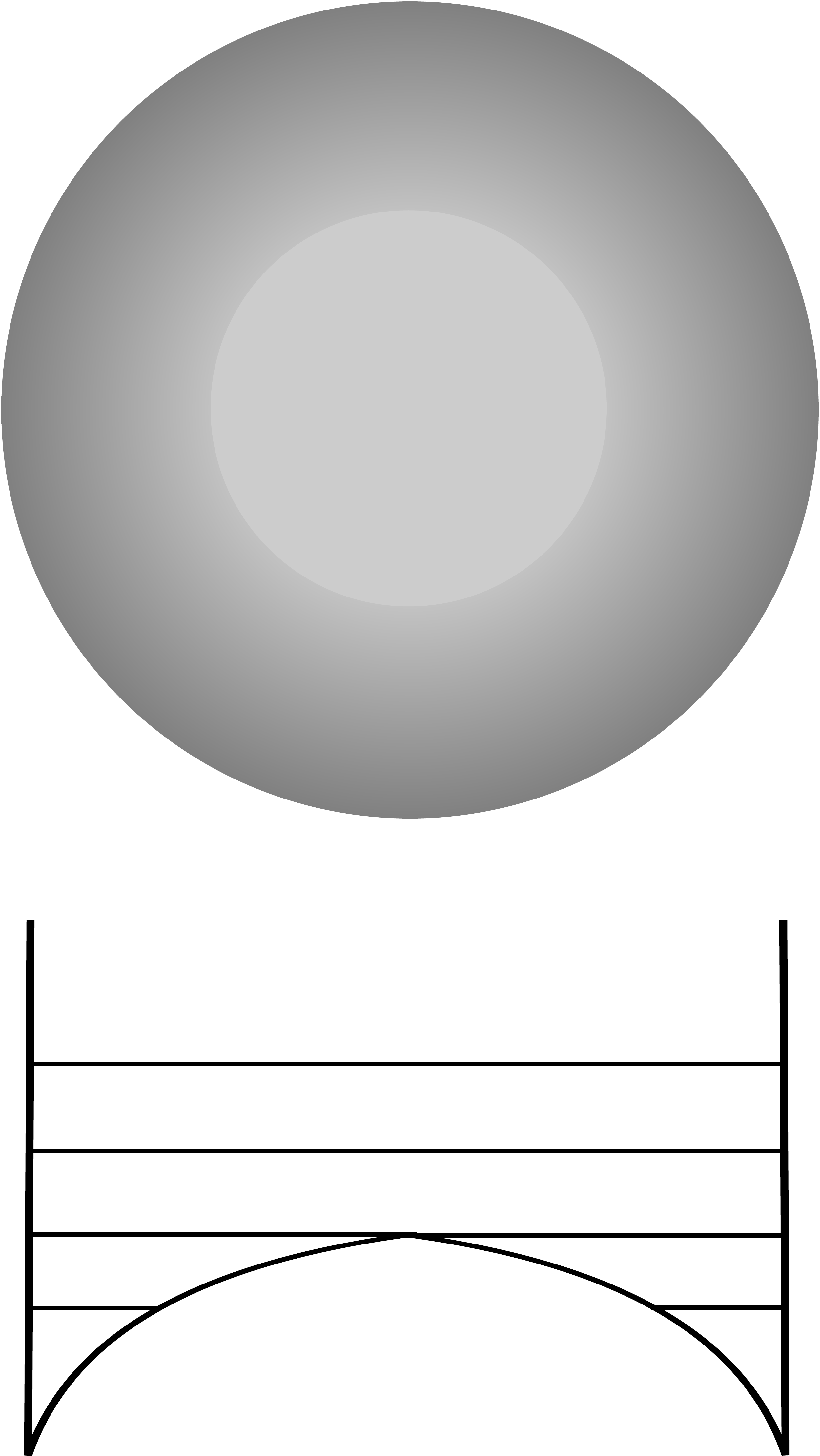}}
	\subfloat{(c)\includegraphics[width=.13\textwidth,valign=t]{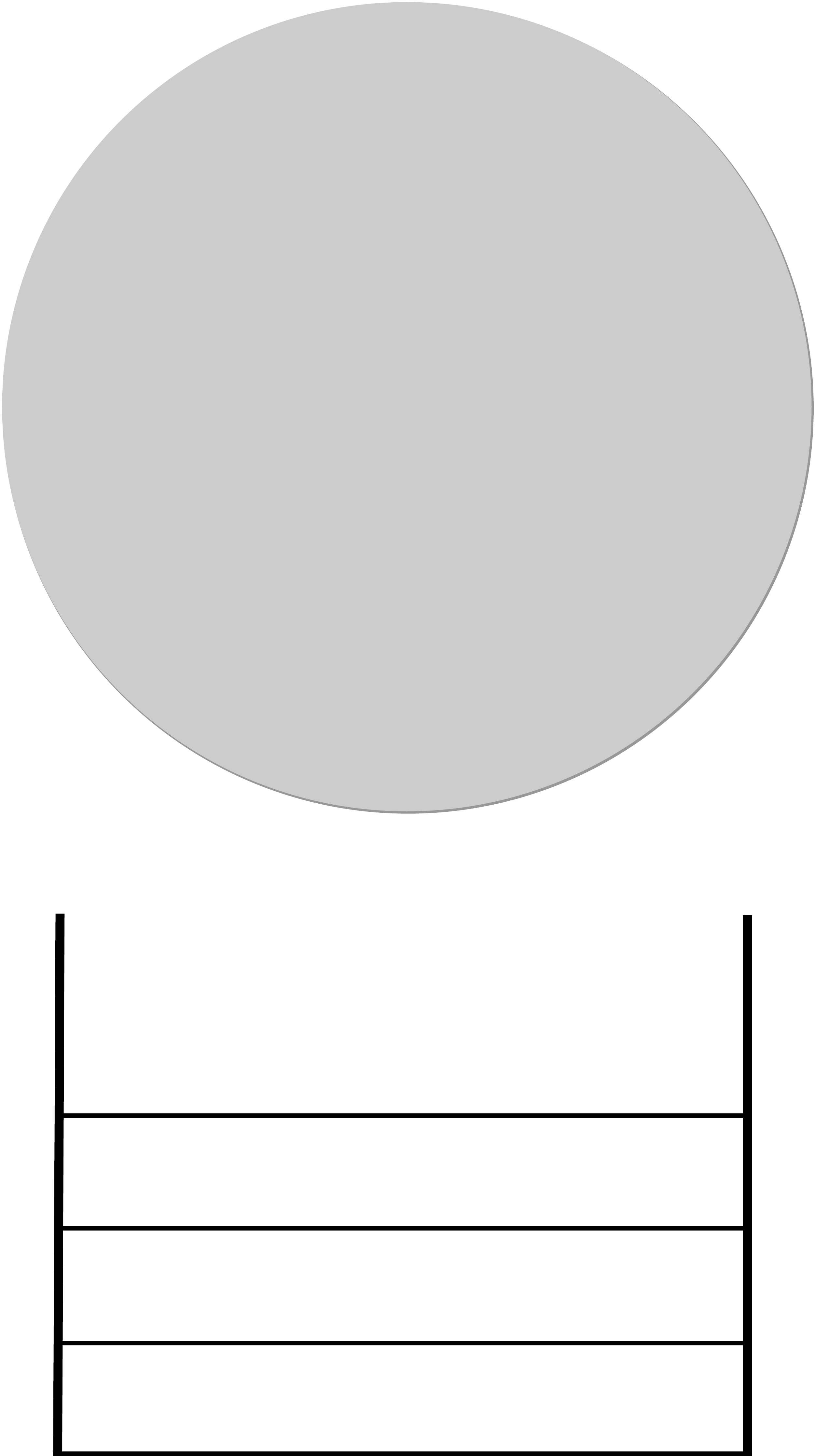}}
	\centering
	\caption{Electron concentration within the nanowire (top) and the corresponding level schematic along the diameter (bottom). Regions of higher concentration correspond to darker shading. a) Regions B and C of the scaling ``phase diagram" Fig. \ref{fig:wire}, where all subbands are confined electrostatically forming an accumulation layer of thickness $D$ near the surface. b) Regions E and F, where the lowest subbands are confined electrostatically, while the top subbands are confined geometrically. c) Regions H and I, where all the subbands are confined geometrically. }
	\label{fig:wiresketch}
\end{figure}

Our results are summarized in Fig. \ref{fig:wire} as a ``phase diagram" in the plane ($\eta$, $R$),  where each ``phase" or region marked by a capital letter denotes a different dependence of the mobility on $R$ and $\eta$ as shown in Table \ref{tab:2}. Just as for quantum wells, many different regions appear due to the interplay between the electrostatic and geometric confinements. The electronic structure of each region is illustrated with a radial level (subband) schematic similar to those in Fig. \ref{fig:well}. One can divide all regions into three groups. In regions D and G the electron gas is strictly one-dimensional (1DEG), i.e. it occupies a single subband in the wire cross-section. In Region A electrons occupy a single radial subband and many azimuthal subbands (2DEG). Finally, in regions B, C, E, F, H, and I, electrons occupy many subbands in both the radial and azimuthal directions and the gas is three-dimensional (3DEG). In order to clarify the meaning of the level schematics,  Fig.  \ref{fig:wiresketch} provides an illustration of the electronic structure in the 3DEG regions. Each top image shows the electron density (shaded regions) in a cross section of the wire while its bottom image shows the corresponding level schematic along the wire diameter.

Let us first concentrate on the 2DEG and 3DEG  regions, where the circumference $2\pi R$ is much larger than the typical electron wavelength $k_F^{-1}$.
This means that we can generalize our results of the quantum well by treating the wire along the $x$ axis as a stripe-like quantum well whose $y$-direction size is $2\pi R$ and 2D concentration $n=\eta/2\pi R$. As a result each of the regions I-VII of Fig. \ref{fig:well} has an analogous region in Fig. \ref{fig:wire} in which the electronic structure near the surface and the mobility are the same upon substituting $n=\eta/2\pi R$ and $L\simeq R$ everywhere. For example, in Region B electrons are confined electrostatically near the wire surface and form an accumulation layer (see Fig. \ref{fig:wiresketch}(a)) whose thickness is given by Eq. (\ref{eq:decay_length}) with $n=\eta/2\pi R$, similar to region II for the quantum well.  By using the correspondence between regions A, C, E, F, H, and I of Fig. \ref{fig:wire} with regions I, III, IV, V, VI, and VII of Fig. \ref{fig:well} we find the wire mobility values for each of these regions as listed in Table \ref{tab:2}.

So far we have shown that in the 2DEG and 3DEG limits of the nanowire, there is a corresponding region in Fig. \ref{fig:well} from which the mobility of the wire may be obtained upon substituting $n=\eta/R$. In regions D and G however, the electron gas in the wire forms a 1DEG for which there is no corresponding region in the quantum well. Let us first concentrate on Region G, where the gas is geometrically confined to a single subband in the plane of its cross-section $(y,z)$ with energy $E_R=\hbar^2/2m^*R^2$ and its wavelength along the wire axis is $k_F^{-1}=\eta^{-1}$. Here $y$ is the azimuthal direction along the wire circumference and $z$ is the radial direction. Due to the roughness, the radius of the wire varies along the wire surface in the $x$ and azimuthal directions by an amount $\delta R=\Delta(k_F^{-1}R/d^2)^{-1/2}$, where $k_F^{-1}R/d^2$ is the typical number of islands over which the electron averages the roughness. These variations lead to a change in the confinement energy that acts as a random scattering potential given by $V=E_R(\delta R/R)$. Using $\hbar/\tau \approx V^2/(\hbar^2k_F^2/2m^*)$ to estimate the scattering rate, we find the mobility in Region G to be 
\begin{equation}\label{eq:mobility 1DEG G}
\mu=\frac{e}{\hbar}\frac{\eta R^7}{\Delta^2d^2}.
\end{equation}
If we increase $R$ so that we enter Region D, the electron gas will instead be confined electrostatically to a single subband of width $D_0=(a_BR/\eta)^{1/3}$. This change amounts to replacing $R$ by $D_0$ in the confinement energy $E_R$. The mobility can thus be obtained by replacing the $R^6$ factor in Eq. (\ref{eq:mobility 1DEG G}) by $D_0^6=a_B^2R^2/\eta^2$ and so the mobility in Region D is given by
\begin{equation}\label{eq:mobility 1DEG D}
\mu=\frac{e}{\hbar}\frac{a_B^2R^2}{\eta^2\Delta^2d^2}(\eta R)=\frac{e}{\hbar}\frac{a_B^2R^3}{\eta\Delta^2d^2}.
\end{equation}
The factor $\eta R$ is unchanged as this came from averaging over an area $k_F^{-1}R$ on the surface and was independent of the confinement in the radial direction. The mobility values given in Eqs. (\ref{eq:mobility 1DEG G}) and (\ref{eq:mobility 1DEG D}) are shown in Tab. \ref{tab:2}.

\begin{table}[h!]
	\caption{\label{tab:2} Mobility $\mu$ in units of $\left(e/\hbar\right)\left(d^4/\Delta^2\right)$ as a function of the linear electron concentration $\eta$ at $d<a_B$ for different regions.}
	\begin{ruledtabular}
		\renewcommand{\arraystretch}{2}
		\begin{center}
			\begin{tabular}{c|c|c}
				A&B&C\\ \hline
				$\quad a_B^2R^2/\eta^2 d^6\quad$&$a_B^{8/5}R^{11/5}/d^6\eta ^{11/5}\quad$&$a_BR/\eta d^3$\\ \hline
				D & E &F\\ \hline
				$a_B^2R^{3}/\eta d^6$&$a_B^{1/2}R^{11/3}/\eta^{11/6}d^6$&$a_B^{1/2}R^{5/3}/\eta^{5/6}d^3$\\ \hline
				G & H & I\\ \hline
				$R^7\eta/d^6$&$R^{14/3}/\eta^{4/3}d^6$&$R^{8/3}/\eta^{1/3}d^3$\\
			\end{tabular}
		\end{center}
	\end{ruledtabular}
\end{table}

We can make the previous discussion more rigorous by considering the scattering rate using Fermi's golden rule. In the 1DEG limit there is only one radial or azimuthal subband occupied so that the scattering rate given by Eq. (\ref{eq:scattering_rate}) then simplifies to
\begin{equation}\label{eq:1DEG scattering rate}
\begin{aligned}
\frac{1}{\tau}=&\frac{2\pi}{\hbar}\frac{1}{R}\sum_{k_y'}\int \frac{dk_x'}{2\pi}<\abs{V(q)}^2>\delta(\varepsilon_F-\varepsilon')\\ =&\frac{2\pi}{\hbar}\frac{1}{R}\int \frac{dk_x'}{2\pi}<\abs{V(q)}^2>\delta(\varepsilon_F-\varepsilon').
\end{aligned}
\end{equation}
Here the marginal one-dimensional screening is ignored and $<\abs{V(q)}^2>$ is defined to be
\begin{equation}\label{eq:matrix element single subband}
<|V(q)|^2>\simeq  \left(\frac{\hbar^2}{m^*Z^3}\right)^2W(q)
\end{equation}
for the gas confined to the lowest radial subband where $W(q)=\Delta^2d^2$ at $k_Fd\ll1$. Setting $Z=R$ in region G and  $Z=D_0$ in Region D, we arrive at the mobilities given by Eqs. (\ref{eq:mobility 1DEG G}) and (\ref{eq:mobility 1DEG D}).

We see in Fig. \ref{fig:wire} that Region G is located at small $R$ and small $\eta$, and extends until the line $R=d$. Beyond this point, the characteristic size of the islands $d$ becomes larger than the radius of the wire $R$ and the model of isotropically distributed islands on the wire surface breaks down.

So far we have dealt with the mobility of quantum wires that are cylindrically symmetric. A stripe-like wire along the x-axis can be made out of a narrow single subband GaAs/AlGaAs quantum well by the etching or split-gate techniques~\cite{Marcus_scalable}. The mobility of such a modulation-doped stripe of 2DEG was calculated~\cite{Ando_2D} for $k_Fd\ll 1$ under the assumption that all scattering happens on the one-dimensional rough $y=0,\,R$ edges of the stripe and that the stripe has many $y$-direction subbands filled. Although our undoped wires studied in regions H and I are different from wires of Ref. \onlinecite{Ando_2D}, they share an important feature with them, i.e., the conduction is determined by the lowest subband. This can be easily understood quasiclassically, as the lowest subband electrons have most of its kinetic energy in the $x$-direction and run approximately parallel to the surfaces or edges, and thus get rarely scattered.

\section{Ballistic-Diffusive Boundary and the Conductance of a Wire with Length $\mathcal{L}$}
\label{sec:conductance}

In the Introduction we explained that due to the 1D nature of the wire the transport properties differ greatly across the different regions of Fig. \ref{fig:wire}. Specifically, in the multisubband regions the wire of characteristic size $R$ undergoes a transition between a ballistic conductor and a diffusive metal as a function of concentration. We will now explain why such a transition occurs, and calculate the conductance $G$ within these regions. 
 
\begin{table}[h!]
	\caption{\label{tab:border} Metal-insulator border $R_{MI}(\eta)$, ballistic-diffusive border $R_{BD}(\eta)$, and the total number of subbands $K_{max}$ for regions G, H, E, and B of Fig. \ref{fig:wire}}
	\begin{ruledtabular}
		\renewcommand{\arraystretch}{2}
		\begin{center}
			\begin{tabular}{c|c|c|c}
				Region&$R_{MI}(\eta)$&$R_{BD}(\eta)$&$K_{max}$\\ \hline
				G&$\eta^{-2/7}R_c(\mathcal{L})^{5/7}$&-&1 \\ \hline
				H &$\eta^{1/14}R_c(\mathcal{L})^{15/14}$& $\eta^{2/13}R_c(\mathcal{L})^{15/13}$ &$(\eta R)^{2/3}$\\ \hline
				E&$a_B^{-3/22}\eta^{5/22}R_c(\mathcal{L})^{15/11}$&$a_B^{-3/7}\eta^{5/7}R_c(\mathcal{L})^{15/7}$&$(\eta R)^{2/3}$\\ \hline
				B & $a_B^{-8/11}\eta^{6/11}R_c(\mathcal{L})^{25/11}$ & $a_B^{-7/9}\eta R_c(\mathcal{L})^{25/9}$&$\eta^{3/5}R^{2/5}a_B^{1/5}$\\ 
			\end{tabular}
		\end{center}
	\end{ruledtabular}
\end{table}
  
Let us first review what we know about the Drude conductance and show where it fails. In Tab. \ref{tab:2} we give the Drude mobility for the various regions of Fig. \ref{fig:wire}. Using these formulas one can calculate the dimensionless Drude conductance $G_D=(h/e)\eta\mu/\mathcal{L}$ per spin for a wire with length $\mathcal{L}$ and linear concentration $\eta$.
  One can then define the metal-insulator transition by the condition $G_D=1$.
  For example, in region H we find that $G_D=R^{14/3}/(\eta^{1/3}R_c(\mathcal{L})^5)$, where $R_c(\mathcal{L})=(\Delta^2d^2\mathcal{L})^{1/5}$ is defined in the Introduction. 
  Using the requirement $G_D=1$, we find the MIT border within Region H to be $R_{MI}(\eta)=\eta^{1/14}R_c(\mathcal{L})^{15/14}$. Similar calculations for regions G, E, and B lead to the $R_{MI}(\eta)$ in Tab. \ref{tab:border}.
  
The dimensionless Drude conductance is valid in all regions where $G_D>1$, but the mean free path $l<\mathcal{L}$. In Region G where there is a single subband occupied, $G_D=1$ and $l=\mathcal{L}$ are the same as long as $\eta a_B>0.5$ where we can safely ignore electron-electron interactions. However in the multisubband regions B, H, and E the conditions are different. This can be understood by realizing that the condition $G_D=1$ is equivalent to $\zeta=\mathcal{L}$, where $\zeta$ is the localization length. When multiple subbands are occupied, $\zeta$ grows larger than $l$, so in the multisubband region we can satisfy the conditions $l\ll\mathcal{L}\ll\zeta$ required for diffusive transport. 

Let us begin with the simplest Region B where all subbands have the same $l$. We define the mean free path as $l=v_F\tau$, where $\tau$ is the relaxation time and $v_F=\hbar k_F/m^*$ is the Fermi velocity. The relaxation time $\tau$ can be calculated from the mobility in Tab. \ref{tab:2}  and we find that in Region B 
\begin{equation}
l=\frac{a_B^{7/5}R^{9/5}}{\Delta^2d^2\eta^{9/5}},
\end{equation}
The border equation is defined by the condition $l=\mathcal{L}$ and is found to be 
\begin{equation}
R_{BD}(\eta)=\frac{\eta R_c(\mathcal{L})^{25/9}}{a_B^{7/9}}
\end{equation}
as shown in Tab. \ref{tab:border}.

In regions E and H there are radial subbands which are geometrically confined. As we showed in Sec. \ref{subsec:physical_well}, subbands that are geometrically confined will have different relaxation times, with higher subbands having smaller relaxation times. As a result $G_D$ in these regions is determined by the lowest radial subbands where $\tau$ and $l$ are largest. In Region E the bottom $M$ radial subbands are confined electrostatically, while the higher subbands are confined geometrically. Similar to Region B the subbands that are electrostatically confined have the same mean free path 
\begin{equation}\label{eq:M subbands mfp}
l_{1-M}=\frac{a_BR^{7/3}}{\Delta^2d^2\eta^{5/3}}.
\end{equation}
These are the lowest subbands that determine $G_D$ and thus setting $l_{1-M}=\mathcal{L}$ leads to $R_{BD}(\eta)$ in Tab. \ref{tab:border}.

Finally, in Region H all radial subbands are geometrically confined and therefore have different mean free paths. The mean free path of the $Nth$ subband $l_N$ is given by 
\begin{equation}\label{eq:Nth subband mfp}
l_N=\frac{R^{13/3}}{\Delta^2d^2\eta^{2/3}N^2}.
\end{equation} 
We see that $l_N\propto N^{-2}$ and the conductance is determined by the lowest radial subband where $N=1$. We can define the diffusive border by the condition that $l_1=\mathcal{L}$, leading to the border equation in Tab. \ref{tab:border}.

Let us now use these results to determine the number $K$ of ballistic subbands at the border. Recall that in the ballistic regions, the dimensionless conductance of the wire is $G_B=K$. At the border $G_B=G_D$, and so using our results of the Drude conductance we can self consistently find the number of ballistic subbands. In Region H, we find $K=k_FR$, in Region E we find that $K=Mk_FR$, and in Region B we find that $K=k_F^2RD$, where $D$ is given by Eq. (\ref{eq:decay_length}) with $n=\eta/R$. These results can be easily understood. For each radial subband there are $k_FR$ azimuthal subbands that contribute equally to the conductance. Then we can generically set $K=(k_FR)K_r$ where $K_r$ will be the number of ballistic radial subbands at the border. In Region H only one radial subband is ballistic, in Region E there are $M$ ballistic radial subbands, and finally in Region B there are $k_FD$ radial subbands which are ballistic. Beyond the border $K_r$ increases as $(l_1/L)^{1/2}$ until $K$ reaches the total number of subbands given in Tab. \ref{tab:border}, where $l_1$ is given by Eq. (\ref{eq:Nth subband mfp}) for $N=1$. The condition $K_r=k_FR$ defines a final border
\begin{equation}
R(\eta)=\eta^{4/11}R_c(\mathcal{L})^{15/11}
\end{equation} 
in regions H and E, beyond which all subbands are ballistic. 

\section{Variable Radius Model of a Nanowire}
\label{sec:FRM}
Previously, we have considered a model of the surface roughness as flat islands of size $d\ll R$ and height $\Delta$ randomly distributed over the surface of the crystal. For the case of the nanowire however, one can imagine another model of roughness in which the radius of the wire varies along its length, but is independent of the azimuthal direction.  We may consider these variations as ring like steps of typical length $d$  and thickness $\Delta$. The step-like nature of the roughness means that we can describe this new model from our old one by restricting the spatial correlator given in Eq. (\ref{eq:exp}) to variations in the $x$-direction. The corresponding Fourier transform of the correlator is then given by
\begin{equation}\label{eq:1Dcorrelator}
W(q_x,q_y)=2\sqrt{2}\pi\Delta^2d(1+q_x^2d^2/2)^{-1}\delta(q_y)
\end{equation}
where $q_x$ is the momentum along the wire's length and $q_y$ is the momentum in the azimuthal direction. We call this model the Variable Radius Model (VRM).

\begin{figure}[t!]
	\includegraphics[width=.47\textwidth]{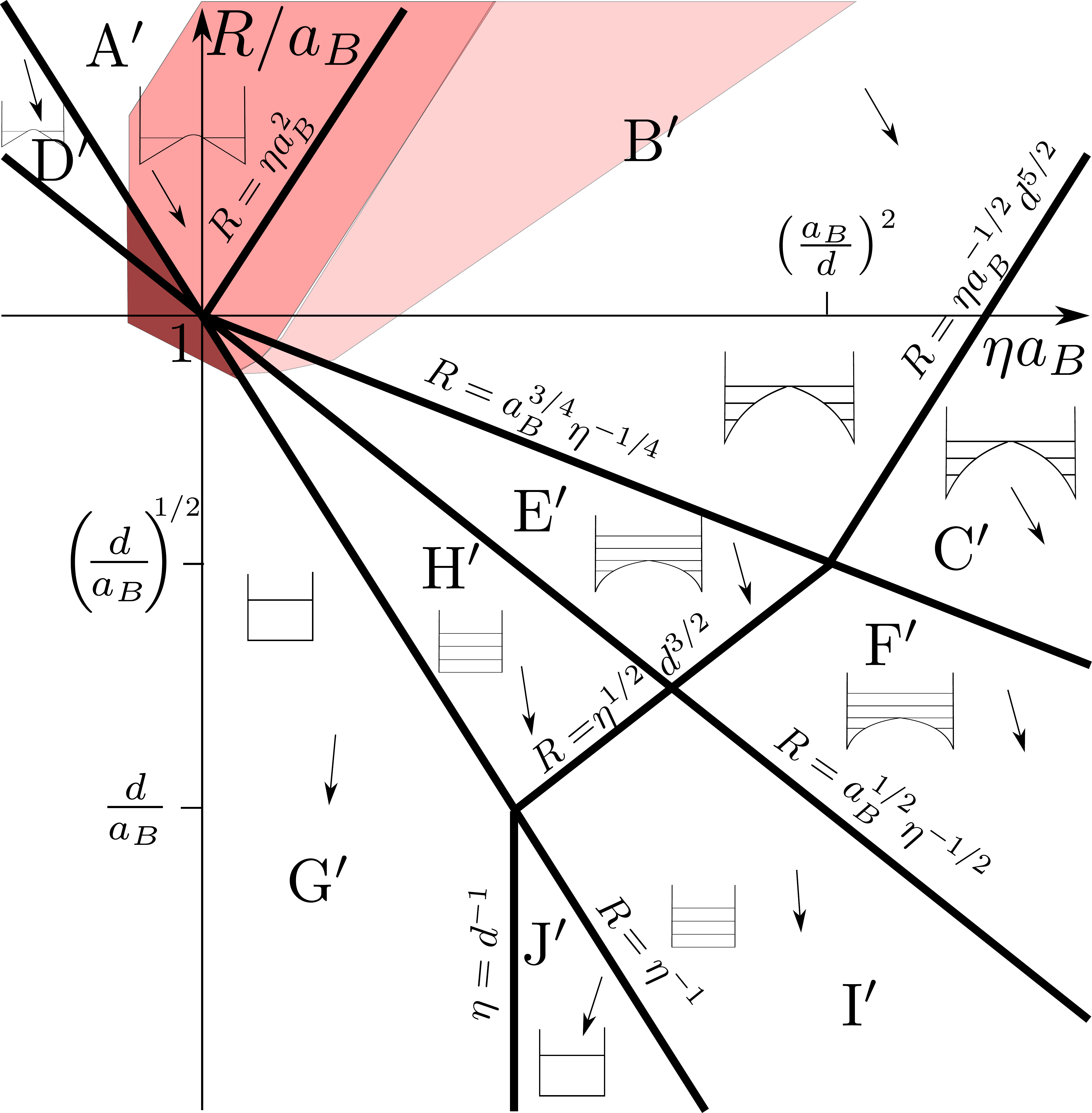}\\
	\caption{The scaling "phase diagram" of roughness limited electron mobility of a quantum wire for the Variable Radius Model (VRM) plotted as a function of radius $R$ and linear electron concentration $\eta$ for $d<a_B$ in the log-log scale. Different "phases" or regions are denoted by capital letters. Mobility expressions corresponding to these regions are given in Table \ref{tab:3}. Region boundaries are given by the equations next to them. The schematic self-consistent electron potential energy profile along the the wire diameter and subbands occupied by electrons are shown for each region. Small arrows show the direction of mobility decrease in each region. All regions and the borders have the same definitions as Fig. \ref{fig:wire}, with the exception of a new region J$'$ that was previously forbidden. The dark red, light red, and pink regions correspond to the single-subband ballistic conductor, many-subband ballistic conductor, and diffusive metal defined by the same conditions as the isotropic model for $\mathcal{L}=a_B^{7/2}\Delta^{-2}d^{-1/2}$. We see that for the same $\mathcal{L}$, the metallic window in the VRM is much smaller than the isotropic model. Electrons are localized in all colorless regions at $T=0$.  }\label{fig:undulatedwire}
\end{figure}

The new phase diagram for the VRM is shown in Fig. \ref{fig:undulatedwire}. It should not be surprising that most of the regions and borders are identical to those in Fig. \ref{fig:wire}, as these are set either by the number of subbands occupied, the type of confinement, or comparison between the island size $d$ and the wavelength $k_F^{-1}$. As none of these properties depend on the details of the correlator, the regions and borders remain the same as Fig. \ref{fig:undulatedwire}. However, there is a new region $J'$ that emerges in Fig. \ref{fig:undulatedwire} that did not appear in Fig. \ref{fig:wire}. This region is the geometrically confined 1DEG under the condition $k_Fd\gg1$. We see that this region occurs in the limit $R\ll d$, which was forbidden for the previous model of roughness. No such restriction is necessary for the VRM, and so the new region emerges.

The mobility of these regions are given in Tab. \ref{tab:3}. We notice immediately that the mobility expressions in regions C$'$, F$'$, and I$'$ are identical to the same lettered regions in Fig. \ref{fig:wire}. The reason is that in these regions, $k_Fd\gg1$, and the scattering is dominated by large angle scattering at the edge of a single island, rather than an effect averaged over many islands. The lack of averaging eliminates the differences between the two models in this region, and so the mobility expressions are the same. When $k_Fd\ll1$, the electrons feel instead an averaged effect, and so we see differences emerge between the two models. The effect of averaging results in a reduction of the scattering rate by the number of scattering centers which are typically seen. In the model considered previously, the variations are two-dimensional and so the electrons average along both the $x$-direction and the azimuthal direction. This leads to an average number of islands that contribute to scattering given by the factor $1/(k_Fd)^2$ in the 2DEG and 3DEG regions, and $R/(k_Fd^2)$ in the 1DEG limit. In the VRM the variations only occur in the $x$-direction and so we do not average in the azimuthal direction. This reduces the number of islands averaged over to be $1/(k_Fd)$ in all regions.
Knowing this, we may easily obtain the new mobilities of most regions by multiplying the expressions in Tab. 2 by the ratio of the new number of islands to the old number of islands. This ratio is $k_Fd$ in the 2DEG and 3DEG and $d/R$ in the 1DEG. The results are shown in Tab. \ref{tab:3}.

\begin{table}[h!]
	\caption{\label{tab:3} Mobility $\mu$ in units of $\left(e/\hbar\right)\left(d^4/\Delta^2\right)$ as a function of the linear electron concentration $\eta$ at $d<a_B$ for different regions of Fig. \ref{fig:undulatedwire}.}
	\begin{ruledtabular}
		\renewcommand{\arraystretch}{2}
		\begin{center}
			\begin{tabular}{c|c|c}
				A$'$&B$'$&C$'$\\ \hline
				$\quad a_B^2R^{3/2}/\eta^{3/2} d^5\quad$&$a_B^{7/5}R^{9/5}/d^5\eta ^{9/5}\quad$&$a_BR/\eta d^3$\\ \hline
				D$'$ & E$'$ &F$'$\\ \hline
				$a_B^2R^{2}/\eta d^5$&$a_B^{1/2}R^{3}/\eta^{3/2}d^5$&$a_B^{1/2}R^{5/3}/\eta^{5/6}d^3$\\ \hline
				G$'$ & H$'$ & I$'$\\ \hline
				$R^6\eta/d^5$&$R^{4}/\eta d^5$&$R^{8/3}/\eta^{1/3}d^3$\\ \hline
				---& --- & J$'$\\ \hline
				---& --- & $\eta^3R^6/d^3$
			\end{tabular}
		\end{center}
	\end{ruledtabular}
\end{table}

While we can understand the changes in mobility in the VRM as due to a difference in averaging, we may also derive these changes from the correlator in Eq. (\ref{eq:1Dcorrelator}). All the differences between the two models occur in the regions where $k_Fd\ll1$, where the correlator is simply $\sqrt{2}\Delta^2d\delta(q_y)$. We see that the major difference from Eq. (\ref{eq:exp}) is that $d^2\rightarrow d$ $\delta(q_y)$, and so it must be true that this difference is what is responsible for the change in the mobility between the two models. Indeed, when calculating the scattering rate, we integrate the correlator over the possible final states $k'$, so that it appears in the scattering rate as a factor $\int d^2k'W(q)$. In our previous model this provided to the scattering rate an overall factor of $k_F^2\Delta^2d^2$ for the 2DEG and 3DEG regions, and $k_F\Delta^2d^2/R$ in the 1DEG. In the VRM the presence of a delta-function for the azimuthal momentum means that these factors change to $k_F\Delta^2d$ in all regions. From here it is clear that the change in the correlator leads to a difference in the mobility between the two models by a factor of $k_Fd$ in 2DEG and 3DEG regions and $d/R$ in the 1DEG regions as we described above.

We have shown that all regions in Fig. \ref{fig:undulatedwire} can be obtained from Fig. \ref{fig:wire} except for the region J$'$. In this region $k_Fd\gg1$, where the scattering rate is determined by large angle scattering. As was discussed in Sec. \ref{subsec:physical_well}, the large angle scattering reduces the correlator, and thus the scattering rate, by a factor of $(k_Fd)^3$ in the denominator. This allowed us to obtain the mobility for $k_Fd\gg1$ from the corresponding region with $k_Fd\ll1$ by multiplying the expression by the factor $(k_Fd)^3$. The same logic may be applied in the VRM, but with a small change. The correlator for the VRM has a different power in the denominator than the previous model. The large angle scattering then reduces the correlator by a factor of $(k_Fd)^2$ in the denominator, rather than $(k_Fd)^3$. This means that we may obtain the mobility of J$'$ from that of G$'$ by multiplying by the factor $(k_Fd)^2=(\eta d)^2$, and this value is shown in Tab. \ref{tab:3}.

The results presented in Sec. \ref{sec:conductance} about the conductance and ballistic-diffusive border can easily be generalized to the VRM model. As the results are quite similar, we do not repeat the discussion here.

\section{Discussion}\label{sec:disc}

Here we would like to estimate the critical value $R_c(\mathcal{L})$ in which the metallic window opens for InAs and InSb nanowires. In order to obtain an accurate estimate of $R_c(\mathcal{L})$, we first need the proper numerical coefficient beyond the scaling approach. Fortunately, the simple single subband structure of regions G and G$'$ allows this number to be determined analytically if we ignore electron-electron interactions. We have calculated these coefficients for a cylindrical wire in Appendix A and found that the mobility in Region G of the isotropic model is 
\begin{equation}\label{eq:mobilityGnumber}
\mu=0.047\frac{e}{\hbar}\frac{\eta R^7}{\Delta^2d^2},
\end{equation}
while for Region G$'$ of the VRM we find the mobility to be
\begin{equation}\label{eq:mobilityGnumberFRM}
\mu=0.017\frac{e}{\hbar}\frac{\eta R^6}{\Delta^2d}.
\end{equation}
$R_c(\mathcal{L})$ is defined to be the radius in which the dimensionless conductance $G=1$. Using Eqs. (\ref{eq:mobilityGnumber}) and (\ref{eq:mobilityGnumberFRM}), and assuming we are on the border $\eta R=1$ between regions G and H (or G$'$ and H$'$), we find the value of $R_c(\mathcal{L})$ in the isotropic roughness model to be 
\begin{equation}\label{eq:Rcisotropic}
R_c(\mathcal{L})=1.8(\Delta^2d^2\mathcal{L})^{1/5},
\end{equation} 
while for the VRM we find
\begin{equation}\label{eq:RcVRM}
R_c(\mathcal{L})=2.8(\Delta^2d\mathcal{L})^{1/4}.
\end{equation}
Now let us see what our theory predicts for a wire with $\mathcal{L}=$ 1 $\mu$m. If we assume that $\Delta=1$ nm and $d=10$ nm, then using Eq. (\ref{eq:Rcisotropic}) we find that $R_c(\mathcal{L})= 18$ nm for the isotropic model, while using Eq. (\ref{eq:RcVRM}) for the VRM we find $R_c(\mathcal{L})=28$ nm. We see that $R_c(\mathcal{L})<a_B$ in both InAs ($a_B\approx34$ nm)\cite{Ford} and in InSb ($a_B=64$ nm)\cite{Pandaya}, so that the ballistic single subband region exists. Recent experiments\cite{Fadaly} have demonstrated ballistic transport in InSb nanowires with $\mathcal{L}\leq1$ $\mu$m and $R$ in the range of $40-50$ nm. These $R$ satisfy the condition  $R_c(\mathcal{L})<R<a_B$ from our estimates, and thus our theory is consistent with their observation of ballistic transport.

In the above estimate we used the condition $G=1$ so that the conductance per spin was $e^2/h$. One could use a different condition in which $R_c(\mathcal{L})$ is defined to be the $R$ such that $l=\mathcal{L}$. This different definition alters $R_c(\mathcal{L})$ by a factor 1.1 in the isotropic model and 1.2 in the VRM, and so our prediction for $R_c(\mathcal{L})$ is only slightly different between the two definitions.

\section{Conclusion}\label{sec:con}

In this paper, we have studied the surface-roughness limited mobility in quantum wells and wires for single-subband and multisubband cases. In these systems, electrons are either electrostatically confined by the surface electric field $E$ or geometrically confined by the surface barriers. The mobility is found to be a function of the electron concentration and well width $L$ or wire radius $R$. Both quantum wells and wires are studied for the exponential model of roughness. For the wires, another model of variable radius (VRM) where there is exponential roughness only in the direction of the wire axis is also discussed. We have presented ``phase diagrams" summarizing the rich collection of mobility scaling regions and found that in quantum wires there exists a critical size $R_c(\mathcal{L})$ so that wires with $R>R_c(\mathcal{L})$ have a window of concentrations where the wire is metallic, while for $R<R_c(\mathcal{L})$ electrons are localized at $T=0$. 

So far we have ignored the spin-orbit coupling of electrons. In InAs and InSb nanowires studied for the purpose of quantum computations \cite{Doh,Frolov,Mourik,Deng_hybrid}, the spin-orbit interaction is quite strong. However, the experimentally relevant Rashba spin-orbit interaction \cite{ilse} just shifts two electron bands of opposite spin polarizations away from each other in the Brillouin zone. Therefore, electrons in each spin polarized band move independently of the other band and the mobility is the same as the case without the spin-orbit coupling.
 
$\phantom{}$
\vspace*{2ex} \par \noindent
{\em Acknowledgments.}

We are grateful to A. Kamenev, V. Pribiag, X. Ying, and K.V. Reich for helpful discussions. Han Fu was supported by the Doctoral Dissertation Fellowship through the University of Minnesota.

\appendix

\section{Coefficients of Mobility for Geometrically Confined 1DEG in Cylindrical Nanowires in Region G and G$'$}\label{App:AppendixB}
In the Discussion, we have used the coefficient of the mobility and thus the mean free path of electrons in narrow nanowires of cylindrical cross-sections at low electron concentrations (Region G of Fig. \ref{fig:wire} and G$'$ of Fig. \ref{fig:undulatedwire}). In this appendix, we derive this coefficient.

For a narrow nanowire at low electron concentrations, electrons occupy only the first subband in the wire cross-section forming a 1DEG which is geometrically confined. If we ignore correlation effects, the wavefunction of the lowest subband in a cylindrical nanowire of radius $R$ is  
\begin{equation}
\xi(r,\phi,x)=\frac{J_0(\nu_0r/R)e^{ikx}}{\sqrt{\pi}RJ_1(\nu_0)}
\end{equation}
where $x$ is directed along the wire axis, $r$ is the distance from the wire center, $\phi$ is the azimuthal angle in the cross section of the wire, $J_0$ and $J_1$ are the zeroth and first order Bessel functions of the first kind, and $\nu_0\approx2.4$ is the first zero of $J_0$.  

It can be easily derived that for a 1DEG, the scattering rate is 
\begin{equation}\label{eq:tau_cylindrical wire}
\frac{1}{\tau}=\frac{2\pi}{\hbar}|V|^2\rho(1-\cos\theta)
\end{equation}
where $|V|$ is the scattering matrix element due to roughness, $\theta=\pi$ is the angle between initial and final electron momenta, $\rho=m^*/2\pi\hbar^2k_F$ is the density of states into which the backscattering can happen, and $k_F$ is the Fermi wavenumber of the 1DEG. For 1D scattering, only backscattering can cause momentum relaxation, and so the angle between the initial and final momenta is $\pi$. 

Similar to Eq. \ref{eq:scattering_matrix element}, according to Ref. \onlinecite{Ando}, one can obtain the scattering potential in the cylindrical geometry to be 
\begin{equation}
V(\phi,z)=\frac{\hbar^2}{2m^*}\Delta(\phi,z)\left.\frac{\partial \xi}{\partial r}\frac{\partial \xi'}{\partial r}\right\vert_{r=R}
\end{equation}
and the scattering matrix element for $R\gg d$ is  
\begin{equation}\label{eq:matrix element cylinder}
<\abs{V(q)}^2>=\frac{\nu_0}{2\pi}\frac{\hbar^4}{m^{*2}R^7}W(q)
\end{equation}
where $q=2k_F$ is the transferred momentum along the wire axis for backscattering of electrons at the Fermi level.

If we combine Eqs. (\ref{eq:tau_cylindrical wire}) and (\ref{eq:matrix element cylinder}), set $k_F=(\pi/2)\eta$ for a 1D gas, and use $k_Fd\ll1$ for the correlator given in Eq. (\ref{eq:exp}), we find the mobility $\mu=e\tau/m^*$ to be
\begin{equation}
\mu=\frac{\pi}{2\nu_0^4}\frac{e}{\hbar}\frac{\eta R^7}{\Delta^2d^2}.
\end{equation}

If instead we consider the VRM model described in Sec. \ref{sec:FRM}, then we use the correlator given in Eq. (\ref{eq:1Dcorrelator}) instead. As a result, the mobility in Region G$'$ in the VRM is
\begin{equation}
\mu=\frac{\pi}{4\sqrt{2}\nu_0^4}\frac{e}{\hbar}\frac{\eta R^6}{\Delta^2d}.
\end{equation}

\section{large angle scattering dominance in scattering rate for quantum well} \label{App:AppendixA}

Here by using Eq. \eqref{eq:scattering_rate} we prove that the scattering rate in $k_Fd\gg1$ regions V and VII is dominated by the large angle scattering, i.e, scattering events with large $q\simeq k_F$. One might expect that because the correlator $W(q)\sim\Delta^2d^/(k_Fd)^3$ for large angle scattering with $q\sim k_F$ is much smaller than that for scattering into small angles with $q\sim 1/d$ by a factor of $(k_Fd)^3$ in the denominator, that the scattering is dominated by the small angle regime. However as we show below, the limited number of final subbands that electrons can scatter into for $q\le 1/d$, the small value of the angular integral $\int d\phi \left(1-\vec{v_{k'}}\cdot\vec{E}\tau_{N'}/\vec{v_k}\cdot\vec{E}\tau_N\right)$, and  in certain cases the smaller $z$-direction momentum of final states $k_z'\ll k_F$ act to suppress the small angle scattering rate so that the scattering is determined by the large angle scattering. We show this below for three cases: $L<d$ (in some part of regions V and VII); $L>d$ and $M/L<1/d$ (for the rest of Region V and some part of Region VII); $L>d$ and $M/L>1/d$ (for the rest of Region VII).

First let us consider the case when $L<d$.  From energy conservation, the total magnitude of the momentum is fixed, and so any difference in magnitude of the in-plane momenta follows from the difference $\abs{k_z-k_{z'}}\sim 1/L$ of their $z$-momentum. When $L<d$, $q\le1/d\ll1/L$ and the scattering happens only within the same subband. This means that the DFR term $\vec{v_{k'}}\cdot\vec{E}\tau_N'/\vec{v_k}\cdot\vec{E}\tau_N$ reduces to the usual $\cos\phi$ for 2D scattering, where $\phi$ is the angle between $\vec{v_{k'}},\,\vec{v_k}$. The final angular integral for the small angle scattering is $\int(1-\cos\phi)d\phi\sim \phi^3\sim(k_Fd)^{-3}$, while it is of order unity for the large angle one. This cancels the advantage of larger $W(q)$ in the small angle scattering. Moreover, the small angle scattering has only one final subband to scatter into while the large angle scattering covers all $k_FL$ subbands. This combined with the small angular integral means that the small angle scattering rate is $k_FL\gg1$ times smaller than that of the large angle when $L<d$.

Now let us look at the second case where $L>d$ and $M/L<1/d$. For simplicity, we focus on the lowest subbands with $k_z<M/L\ll k_F$ as these dominate the conductivity in regions V and VII. In the limit $L>d$ and $M/L<1/d$, there will always exist $L/d>M$ subbands with $k_z'< 1/d$ so that the scattering now involves intersubband scattering. As a result the DFR term is not reduced to $\cos\phi$ and the term $\left(1-\vec{v_{k'}}\cdot\vec{E}\tau_{N'}/\vec{v_k}\cdot\vec{E}\tau_N\right)$ is of order unity instead of being infinitesimal for small $q$ and thus small $\phi$. The angular integral of this term would just give $1/(k_Fd)$ from the small angle $\int d\phi\simeq 1/dk_F$ and does not compensate the $(k_Fd)^3$ reduction of the correlator.  However, we must consider the importance of $k_z'$ in the scattering matrix element according to Eq. \eqref{eq:scattering_matrix element}. For the small angle scattering $k_z'<1/d$, while for the large angle regime $k_z'\sim k_F$. This gives an extra factor $1/(k_Fd)^2$ to $|V(q)|^2$ in small angle regime relative to the large angle scattering. This additional factor combined with the small angular integral compensates the $1/(k_Fd)^3$ reduction of the correlator. Considering also the accessible number of final subbands $L/d$ in the small angle limit is $k_Fd$ times smaller the $k_FL$ available subbands for large angle scattering, we find that the small angle scattering rate is $k_Fd$ times smaller than that of the large angle.

Finally we must consider the intermediate case when $L>d$ and $M/L>1/d$. For small angle scattering the number of subbands $L/d$ that may be scattered into is small, and so we expect that the DFR term is near the 2D limit $\cos\phi$. Expanding around this value, we find that the DFR term is approximately
\begin{equation}\label{eq:DFR}
\frac{\vec{v_{k'}}\cdot\vec{E}\tau_N'}{\vec{v_k}\cdot\vec{E}\tau_N}=\cos\phi(1-\frac{\delta v_k}{v_k}-\frac{\delta\tau_N}{\tau_N}),
\end{equation}
where $\delta v_k=\abs{v_k-v_k'}$ and $\delta\tau_N-\abs{\tau_N-\tau_N'}$.
Let us examine these correction terms, beginning with $\delta v_k/v_k$. The allowed difference in $k$ and $k'$ is $1/d$ for the small angle scattering. Since $k\simeq k_F$ for lowest subbands, the velocity difference ratio $ |v_{k'}-v_k|/v_k=|k'- k|/k$ is then $1/k_Fd$. In considering the other correction term $\delta\tau_N/\tau_N$,
let us assume that the scattering rate of each subband is always determined by their large angle scattering into typical subbands and show that this assumption self-consistent. With this assumption the difference in relaxation times $\delta\tau_N$ is solely caused by the different $z$-direction momenta and subband widths as seen from Eq. \eqref{eq:scattering_matrix element}. Again we focus on the the lowest $M$ subbands as these determine the conductivity. For the bottommost subbands, all subbands within $q\sim 1/d$ are electrostatically confined and $\delta\tau_N=0$ as $k_z'^2/Z'=k_z^2/Z=n/a_B$ (see Eq. \eqref{eq:scattering_matrix element}) For the higher subbands with $k_z\sim M/L$, there are bands within $q\sim1/d$ which are instead geometrically confined and the correction is non-vanishing. Indeed, we find that $\delta\tau_N\sim\tau_N\delta k_z/k_z$ and so the correction is given by $(1/d)/(M/L)$. We find then that the leading contribution to the DFR term in Eq. (\ref{eq:DFR}) in the small angle regime is approximately $1-(1/d)/(M/L)$, where we have used the fact that $1/(k_Fd)\ll L/Md$ in the limits being considered.

Using the DFR term above, the angular integral now gives a factor $(1/k_Fd)(1/d)/(M/L)$ to the scattering rate, while the integral is of order unity for the large angle limit. Combined with the fact that the final state in the small angle regime has $k_z'^2/Z'\simeq (M/L)^2/L$,  we find that these terms give an extra factor $(Md/L)/(k_Fd)^3$ compared to the same terms for the large angle limit. We see then that there is a factor of $1/(k_Fd)^3$ term that compensates the suppression of the correlator in the large angle limit. Adding the fact that the small angle scattering can only scatter into $L/d\ll k_FL$, we find that the ratio of scattering rates in the small and large angle regimes is $M/(k_FL)\ll1$ and indeed the large angle limit dominates.

\bibliography{wire}
\end{document}